\documentclass{aa}
\usepackage{graphics,psfig}
\begin{document}

   \thesaurus{11     
              (11.05.2; 
11.19.2; 
11.19.7; 
11.19.6)} 

\title{Box- and peanut-shaped bulges
\thanks{
Partly based on observations collected at ESO/La Silla (Chile), DSAZ/Calar Alto
(Spain), and Lowell Observatory/Flagstaff (AZ/USA).}
}

\subtitle{I. Statistics}
\author{R. L\"utticke \and R.-J. Dettmar \and M. Pohlen}

\institute{Astronomisches Institut, Ruhr-Universit\"at Bochum,
D-44780 Bochum, Germany\\
email: luett@astro.ruhr-uni-bochum.de
}

\date{Received 11 May 2000; accepted 14 June 2000}

\titlerunning{Box- and peanut-shaped bulges. I}
\maketitle

\begin{abstract}

We present a classification for bulges of a complete sample of $\sim$\,1350
edge-on disk galaxies derived
from the RC3 (\emph {Third Reference Catalogue of Bright Galaxies},
de Vaucouleurs et al. \cite{rc3}). 
A visual classification of the bulges using the Digitized Sky Survey (DSS)
in three types of b/p bulges or as an elliptical type
is presented and supported by CCD images.
NIR observations reveal that dust extinction does almost not
influence the shape of bulges.
There
is no substantial difference between the shape of bulges in the optical
and in the NIR.
Our analysis reveals that 45\,\% of all
bulges are box- and peanut-shaped (b/p). The frequency of b/p bulges for all
morphological types from S0 to Sd is $>$ 40\,\%.
In particular, this is for the first time that such a large frequency of
b/p bulges is reported for galaxies as late as Sd.
The fraction of the observed 
b/p bulges is large enough to explain the b/p bulges
by bars. 


\end{abstract}

\section{Introduction}
Several theories of bulge formation are currently discussed: 
primordial scenarios  (monolithic collapse: 
Eggen et al. \cite{ELS} (ELS); clumpy collapse:
Kauffmann et al. \cite{kau}, Baugh et al. \cite{bau};
inside-out formation: van den Bosch \cite{bosch}, Kepner \cite{kep}), 
secular evolution scenarios (merger: Wyse et al. \cite{wys} and references
therein, dynamical evolution due to gravitational
instabilities, such as bars: Combes et al. \cite{com90}), and
combinations of both (Combes \cite{com99}). 
Therefore statistics of bulges are needed to test
the proposed dynamical processes of bulge formation.
Statistics of b/p bulges are of particular interest to demonstrate
the importance of these structures for disk galaxies and to 
point out the relevant evolution scenarios of bulges in general. 
From the frequency
of b/p bulges the likeliness of their formation process can be
estimated.

External cylindrically symmetric torques (May et al. \cite{may}) or mergers
of two disk galaxies (Binney  \& Petrou \cite{bin}, Rowley \cite{row})
as origins of b/p bulges require very special conditions (Bureau \cite{bur}).
Therefore such evolutionary scenarios can explain only a very low
frequency of b/p bulges. Accretion of satellite galaxies 
is a formation process of b/p bulges
(Binney  \& Petrou \cite{bin}, Whitmore \& Bell \cite{whi}), which could
produce a higher frequency of b/p bulges. However, an oblique impact angle
of the satellite is needed for the formation of a b/p bulge and
a massive accretion event would disrupt the stellar disk
(Barnes \cite{bar}, Hernquist \cite{her}). The origin of b/p bulges
by bars was first noticed by Combes \& Sanders (\cite{com81}).
Combes et al. (\cite{com90}) and Raha et al. (\cite{rah})
substantially revived this idea. Comparisons between the frequency of
barred galaxies and galaxies with a b/p bulge can test the probability
of evolutionary scenarios of b/p bulges based on bars.

The estimated frequency of galaxies with a b/p bulge has steadily
increased in former statistics
from 1.2\,\% (Jarvis \cite{jar}), over 13\,\% (de Souza \&
dos Anjos \cite{sa87}, hereafter SA87) and 20\,\% (Shaw \cite{sha87}), 
up to 45\,\%
(Dettmar \& Barteldrees \cite{det88}). This rise results from
differences in sample selection, sample size, detection method, and criteria to
identify b/p bulges.
However, the later three statistical studies
have shown that b/p bulges are not really as peculiar
as they were supposed to be in the past.
Therefore very common processes are required to
explain their high frequency.
No complete statistics
and list of galaxies with b/p bulges, based on observations,
have been published in the last ten years.
However, the knowledge about galactic evolution
would greatly benefit from
a reliable value for the frequency of b/p bulges 
by the determination of
the most likely scenario for the formation of b/p bulges and
subsequent conclusions on the evolution of bulges in general. 
A complete list of b/p bulges is 
a base for a
detailed analysis of these structures.

In {\it Sect. 2} we present our investigated sample of edge-on disk
galaxies, and our classification of the bulge shape  
is explained in {\it Sect. 3}.
The results of our statistics and 
of additional follow-up CCD observations in the optical and
in the NIR are given in {\it Sect. 4}.
{\it Sect. 5} compares our findings to former statistics.
In {\it Sect. 6} we discuss our results and finally, in
{\it Sect. 7} we give the 
conclusions from our statistical analysis of the sample.

\section{Sample selection}

\subsection{Former samples}
The first search for b/p bulges was conducted by Jarvis (\cite{jar}).
By visual inspection of all fields of the ESO/SERC $J$ sky survey
he found therein 30 b/p bulges
and listed additionally 11 galaxies with b/p bulges
from the literature. Together with a control sample
he estimated that about 1.2\,\% of all disk
galaxies possess b/p bulges.

Shaw (\cite{sha87}) inspected visually the ESO/SERC $J$ and
Palomar Observatory/National Geographical Survey (POSS) $B$- or $R$-band
scanned survey plates.
He derived a frequency 
of b/p bulges of 20($\pm4$)\,\% as a lower limit
detecting 23
b/p bulges in a sample of 117 disk galaxies. The sample was mainly selected
from the RC2 (\emph {Second Reference Catalogue of Bright Galaxies},
de Vaucouleurs et al. \cite{rc2}) and included disk galaxies 
with a diameter $log D_{25} \ge 1.55$ ($D_{25}\!\sim\!3.5\arcmin$)  
at the 25 \nolinebreak[3]
($B$) mag arcsec$^{-2}$ isophotal level and a ``sufficiently'' edge-on
aspect.  

Until today the largest investigated sample of 555 galaxies was presented by
SA87. Their list was extracted from several catalogues and contained all disk
galaxies  with restricted total $B$ magnitude ($B_{\rm T} < 13.2$ mag)
and axis ratio ($b/a < 0.5$). 
The galaxies were inspected on film copies of the
ESO Quick Blue survey or the POSS prints by means of a microscope. 
A ``substantial
number'' of candidates
were scanned using the PDS microdensitometer at ESO and checked with an 
image processing system (IHAP).
In this way SA87 found 74 b/p galaxies as a lower
limit. This is 13\,\% of their total sample.

A much larger frequency of b/p bulges was proposed after the investigation
of 73 galaxies obtained with CCD surface photometry
(Dettmar \& Barteldrees \cite{det88}, Dettmar \cite{det89}).
That sample gave
a frequency of $45(\pm8$)\,\%
and Dettmar (\cite{det96}) concluded from this sample a lower limit of
the total frequency of
b/p bulges of 35\,\%.
However, only 15 objects
with b/p bulge from this sample were listed by name in a paper of Shaw et al. 
(\cite{sdb})
together with other known b/p bulges from the literature.

\subsection{New sample selection}
The former statistics in mind a new one should contain a large number
of galaxies, be complete, and based on
a detection method for the b/p bulges with reasonable accuracy.
In the new statistics presented here the completeness with regard to the size
of a galaxy is reached by the
selection of all disk galaxies ($-3.5 < T < 9.5$)
out of an electronic version of the RC3
with diameters
larger than 2$\arcmin$ ($\log D_{25} \ge  1.3$). 
For disk galaxies without known $T$ parameter it 
is determined separately (see notes to Table \ref{listrc3}).
Extracting the sample from one catalogue is preferable to the method of
SA87 since the parameters differ among the 
catalogues and an exact selection criterion is impossible. Similar to
Shaw (\cite{sha87}) the selected
sample is diameter limited, but compared to his sample
the limit is nearly 2
times smaller.
Additionally, the use of the newer
catalogue (RC3 instead of RC2) leads to a much larger number
of investigated galaxies. Restricting the sample in diameter and not in
$B_{\rm T}$ magnitude is preferred because
the RC3 attempts to be complete for
galaxies with an apparent
diameter larger than 1$\arcmin$  at the $D_{25}$ isophotal level and in addition,
the $B_{\rm T}$ magnitude is not even listed for all galaxies larger than 2$\arcmin$.
Therefore a magnitude limit would not result in a
complete sample. Furthermore, a magnitude limit would prefer the
selection of early type galaxies. 

Since b/p bulges can only be observed at inclinations down to
$i\sim$\,75$^{\circ}$ (Shaw et al. \cite{sdb}),
in a first step all face-on galaxies are excluded using the axis ratio
according to SA87. S0 galaxies ($-3.5 < T \leq -0.5$)
with $\log R_{25} \ge 0.30$ 
and other disk galaxies with
$\log R_{25} \ge 0.35$ are included ($R_{25}\!=\!\frac{a}{b}$).
These different limits are
used since the transformation of the axis ratio into
the inclination angle depends on the morphological type
(Bottinelli et al. \cite{bot}, Guthrie \cite{gut}). With different formulae
(the simplest is cos $i\!=\!\frac{b}{a}$; see references in Guthrie \cite{gut})
this restriction results
for disk galaxies in a limit for the inclination angle $i$ between 60$^{\circ}$
and
70$^{\circ}$ degree. This limit ensures most likely
a detection of all observable b/p bulges.
The final sample meeting these selection criteria
contains 1343 galaxies.

The preferable method to classify a b/p bulge is the use of CCD images, but
for the whole sample the amount of observing time would be unreasonably large.
Therefore all galaxies are inspected using the Digitized Sky Survey (DSS)
\footnote{http://arch-http.hq.eso.org/cgi-bin/dss}.
This survey is complete over the
whole sky and an investigation of the images of the galaxies
with data analysis software is possible.
The DSS is  based on photographic surveys 
of the northern
POSS $E$ plates ($R$-band, m$_{\rm lim}\!=$ 20.0 mag),
the southern SERC $J$ plates (equivalent to $B$-band, m$_{\rm lim}\!=$ 23.0 mag),
and the southern Galactic plane
SERC $V$ plates ($V$-band, m$_{\rm lim}\!=$ 14.0 mag) (McLean \cite{mcl}) and
has a scale of $1.7\arcsec {\rm pixel}^{-1}$. 
Galaxies which are saturated in their central regions are checked
(if possible) with the ESO Lauberts-Valentijn Archive 
(Lauberts \& Valentijn \cite{lv}) kindly made available by ESO
\footnote{http://archive.eso.org/wdb/wdb/eso/esolv/form}. 
While the images within the
ESO Lauberts-Valentijn Archive are not saturated and have a better scale
($1.35\arcsec$ pixel${^-1}$), they
have a lower signal-to-noise ratio compared to the images of the DSS.

\section{Classification of box/peanut bulges}

Shaw (\cite{sha93a}) and L\"utticke (\cite{lue96}) 
derive objective parameters to characterize
quantitatively  b/p structures by the measurement of the excess luminosity and
the total or maximal
fraction of the b/p distortion
(depending on the radial distance from the center)
in relation to the observed bulge luminosity.
However, the application of such a 
classification method is unpractical for
a large sample of galaxies and strongly depends on the modelled
luminosity distribution for the disk and the elliptical part of the bulge.

The $a_4$ isophote shape parameter is used by Combes et al. (\cite{com90}), 
Shaw (\cite{sha93a}), L\"utticke (\cite{lue96}), and
Merrifield \& Kuijken (\cite{mer}) to quantify 
the degree of boxiness of bulges by measuring the deviations
from perfect ellipses (e.g. Bender \& M\"ollenhoff \cite{ben}).  
However, it is problematic that the determination of the extreme value
of $a_4$ depends on the fitted region of the bulge. 
Fitting only the inner parts of bulges results in some galaxies
in the undetection of the boxiness of the bulge which is
most prominent in the outer parts.  
For instance the clear boxy bulge of NGC 1055
(Shaw \cite{sha93b}, L\"utticke et al. \cite{lue2000b}, hereafter Paper III)
is undetected by Merrifield \& Kuijken (\cite{mer}) using the $a_4$
parameter.  
An additional disadvantage of the determination of this
parameter is the influence  of the masked stars in the foreground,
dust, bars, and the extreme nature of b/p
isophotal distortions for the ellipse fitting. 
Therefore the region of the galaxy which is used for
the fitting is important.
Shaw (\cite{sha93a}) uses the whole galaxy, L\"utticke (\cite{lue96})
the disk subtracted galaxy,
and Merrifield \& Kuijken (\cite{mer}) mask
out a wedge-shaped region of each image
within 12 degrees of the disk major axis and only fit
``on the side of the galaxy
where the disk projects behind the bulge''.

These disadvantages lead
to the fact that the $a_4$ parameter is not suitable for uniform classification
of a large sample of bulges. Additionally, the resolution of the DSS
images is too low to determine the $a_4$ parameter for the whole sample.
Therefore the bulges of the investigated sample are classified by their degree
of b/p shape derived by visual inspection from contour plots of the 
galaxy images.

Reshetnikov \& Combes (\cite{res}) use similar arguments to
choose a ``straightforward procedure of eyeball estimation'' with isophotal
maps from the DSS instead of
objective criteria for the detection and classification of warped disks.

The bulges are divided in three types of b/p bulges ({\bf 1 - 3}),
elliptical bulges, 
and unclassifiable bulges.

\begin{description}
\item \hspace*{4mm} {\bf 1 :} peanut-shaped bulge
\item \hspace*{4mm} {\bf 2 :} box-shaped bulge
\item \hspace*{4mm} {\bf 3 :} bulge is close to box-shaped, not elliptical
\item \hspace*{4mm} {\bf 4 :} elliptical bulge
\item \hspace*{4mm} {\bf 5 :} unclassifiable bulge
\end{description}

\begin{figure}[htbp]
\centering
\psfig{figure=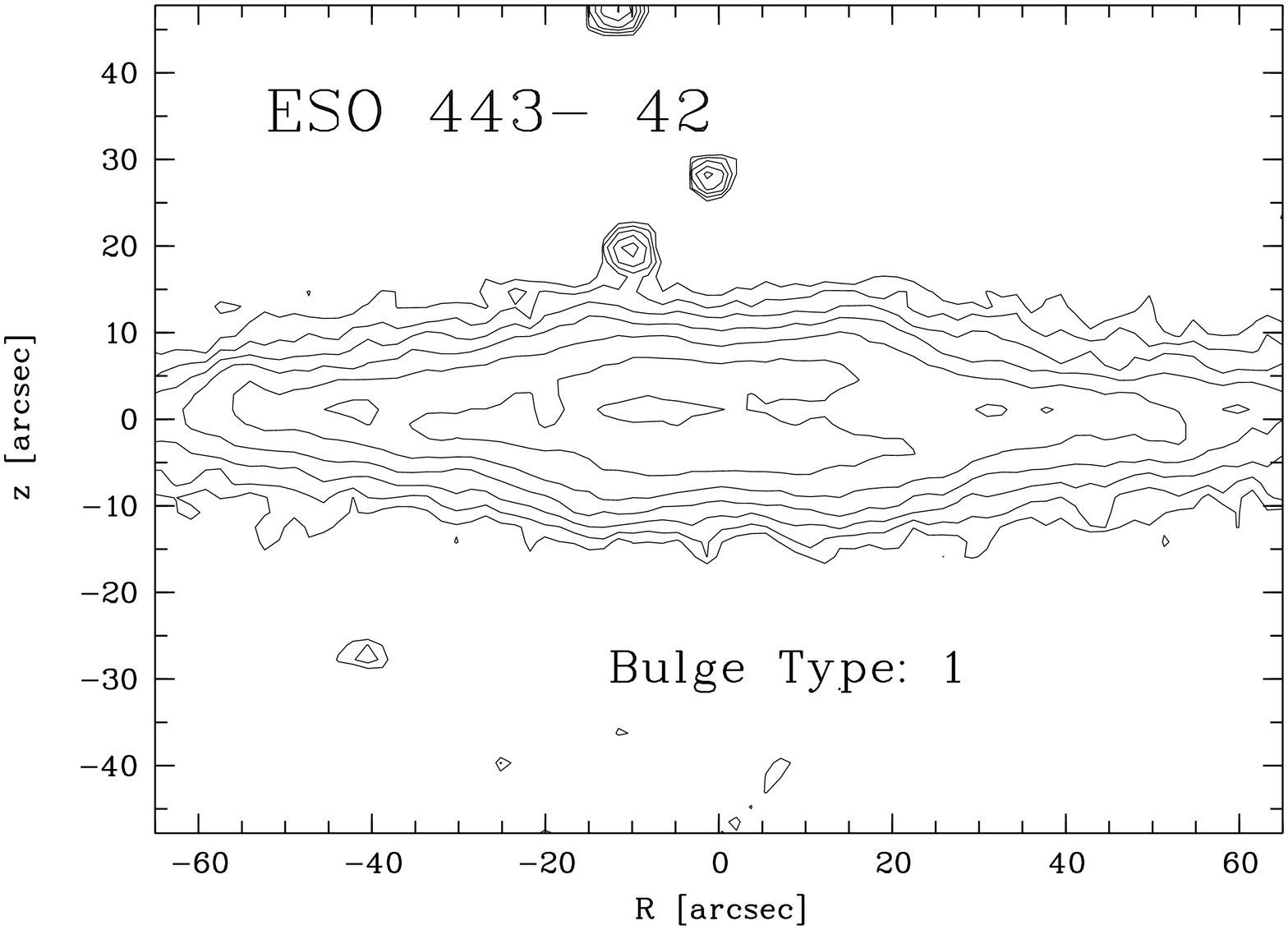,width=8.15cm,angle=270}
\psfig{figure=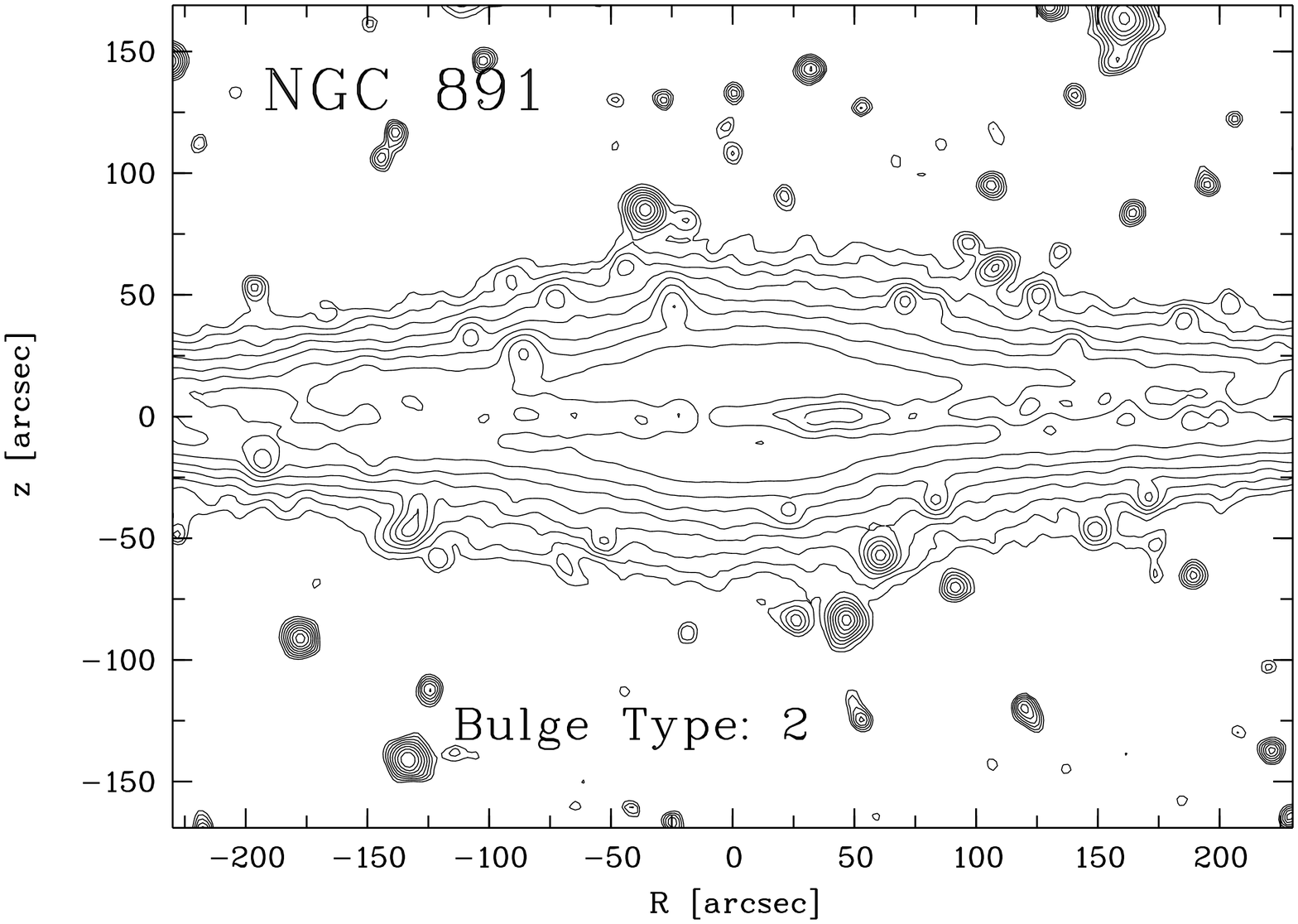,width=8.15cm,angle=270}
\psfig{figure=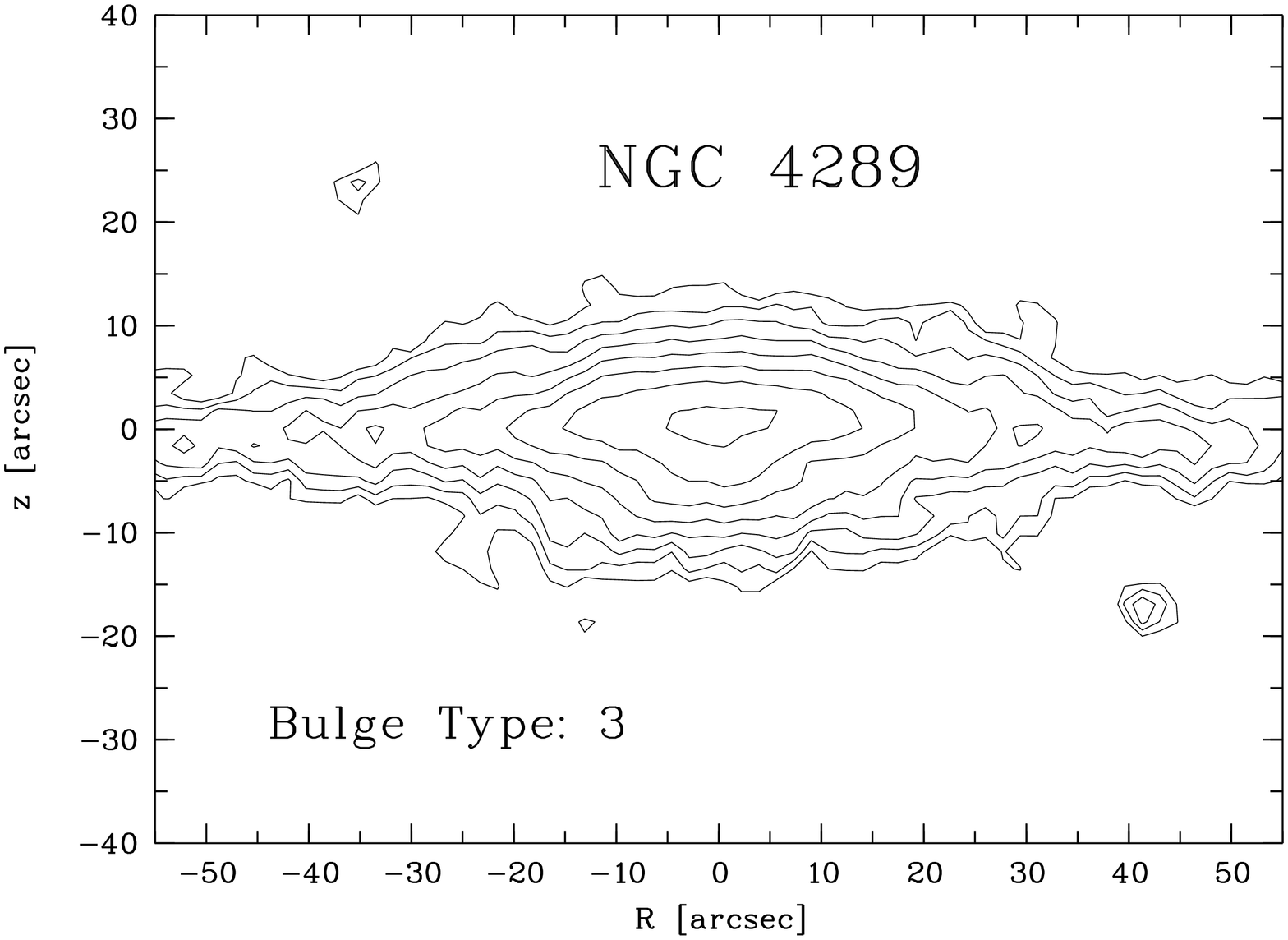,width=8.15cm,angle=270}
\psfig{figure=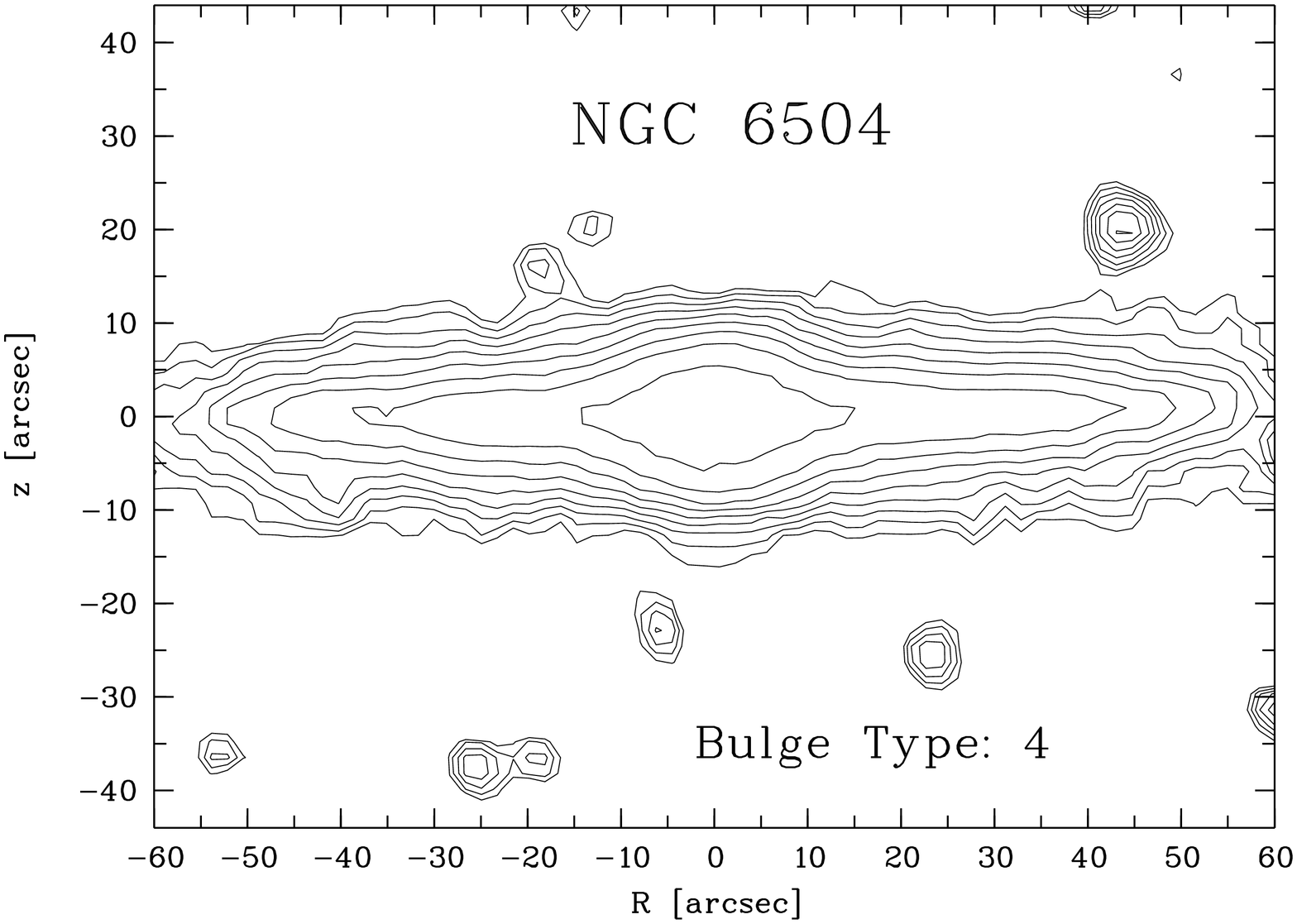,width=8.15cm,angle=270}
\caption{Examples for the different bulge types from the DSS.
Isophotes are logarithmically equidistant.}
\label{dss}
\end{figure}

Type {\bf 1} bulges are described by a depression along the minor
axis on both sides of the main axis. In this way the bulge looks like a peanut
(Fig. \ref{dss}, top).
The depth of the depression can be used as a characteristic parameter
for these peanut-shaped bulges (L\"utticke \cite{lue99}).
The box-shaped bulges (type {\bf 2})
are defined by
isophotes parallel to the major axis. 
Therefore the bulge appears like a box
(Fig. \ref{dss}, second image).
Some bulges of this class have a very prominent box form 
(Fig. \ref{n1886}, top). 
They could be type {\bf 1}, but due to 
low resolution of the images it is not possible to see the eventual
depression along the minor axis. 
These bulges are called {\bf 2+}.     
The same class is used for bulges, 
which are on one side boxy and on the other peanut-shaped.
Type {\bf 3} (Fig. \ref{dss}, third image)
is less  well defined.  These bulges possess a
general flattening of the isophotes parallel to the major axis which is 
less pronounced as the flattening of type {\bf 2}. However, this flattening
differentiates these bulges from the purely elliptical bulges of type {\bf 4}
(Fig. \ref{dss}, bottom). The limits between the classes are not sharp:
 obvious  between {\bf 1} and {\bf 2}, clear
between {\bf 2} and {\bf 3},
but sometimes indistinct between {\bf 3} and {\bf 4}. The
classification of bulges can therefore in some cases
be ambiguous, but a check of the whole
sample (two independent classifications by the author and one 
of the co-authors) 
shows that more than 90\,\% of the bulges are well defined by
one of the different bulge types.

Several reasons are differentiated why bulges are not classifiable:
\begin{description}
\item \hspace*{1mm} {\bf 5.1 :} Inclination is too far away from edge-on.
\item \hspace*{1mm} {\bf 5.2 :} Stars in the foreground projected onto the bulge.
\item \hspace*{1mm} {\bf 5.3 :} Galaxy is strongly perturbed.
\item \hspace*{1mm} {\bf 5.4 :} Dust conceals the shape of the bulge.
\item \hspace*{1mm} {\bf 5.5 :} Bulge is too small.
\item \hspace*{1mm} {\bf 5.6 :} Signal-to-noise ratio of the image is too low.
\end{description}

For galaxies which meet more than one of the above reasons
only the most important one is listed
(Table \ref{listrc3} and \ref{listf}).
The classification for all galaxies in the sample with some minor changes
after inspection of CCD images (Sect. 4.2) are given in Table \ref{listrc3}.

\section{Results}

\subsection{Statistics}
About 3/4 of the Sdm and Sm galaxies are not at all classifiable and  the
rest is highly uncertain, because the bulges of these galaxies --- if they have
any --- are very small and
faint and the influence of dust is very strong.
Other structures like bars or star forming regions are more dominant.
Therefore the selected sample is reduced to disk galaxies ranging from S0-Sd
($-3.5 < T < 7.5$) resulting in  1224 galaxies which will be
investigated in the following.

40.0\,\% of these bulges are unclassifiable, almost 3/4 of them
due to an inclination which is too far from edge-on.
This is an expected effect since the selection criterion of our sample
for the inclination is a lower limit (see above).
Only a few galaxies with a classifiable bulge in the sample have an axis
ratio which is near to the sample limit of $\log R_{25}\!=\!0.30$ (S0) and
$\log R_{25}\!=\!0.35$ (Sa - Sd), respectively. 
This shows that the limits are well chosen.
There are only  two  peculiar edge-on galaxies known
(with $\log D_{25} > 2\arcmin$)  which have  
classifiable bulges and are not in our sample due to their low axis ratio
(IC 2560 and NGC 7123).
Therefore our list of galaxies with b/p bulges and a diameter
larger than 2$\arcmin$ is almost complete and the number of
non-included b/p bulges is small.

Stars in the foreground (type 5.2) are responsible for 13\,\%
of unclassifiable bulges. These stars can influence the projected
light from the galaxy in a way that the shape of the bulge cannot
be identified.
Dust lanes (5.4) prevent classifications in the same way, but
only for a few galaxies, in which the dust lane
has a large extension in comparison
with the bulge size (see also Sect. 4.3). Their
fraction decreases
from late types and is equal to zero for galaxies earlier than Sbc.
Scd and Sd galaxies have too small bulges (5.5) for classification, if
their diameter is around the limit of 2$\arcmin$
and their inclination is near to $90^{\circ}$.
At this orientation the measured diameter ($D_{25}$) and
therefore the ratio of disk to bulge length is maximal
due to optical depth effects
(Xilouris et al. \cite{xil}).
The result is that
galaxies of the same morphological type and diameter ($D_{25}$)
have bulges of different sizes
which depend on the inclination angle.
However, the fraction of galaxies in the sample having such small bulges
is negligible (Table \ref{uncla}).
In a few cases galaxies are in the southern Galactic plane (5.6),
where the DSS has a higher surface brightness limit due to the use
of the SERC $V$ plates (see above).
Therefore the images of the galaxies in this region have frequently a very small
signal-to-noise ratio
and the bulges appear to be too faint for classification.
Perturbations (5.3) by interactions are only in a very few cases the reason for
bulges being not classifiable.

\begin{table}[hbtp]
\caption{Reasons for the unclassifiability of bulges}
\begin{center}
\begin{tabular}{l|cccccc}
bulge type & {\bf 5.1} & {\bf 5.2} & {\bf 5.3} & {\bf 5.4} & 
{\bf 5.5} & {\bf 5.6} \\
\hline
frequency & 74\,\% & 13\,\% & 2\,\% & 4\,\% & 4\,\% & 3\,\% \\
\end{tabular}
\end{center}
\label{uncla}
\end{table}

\begin{table}[hbtp]
\caption{Classifiable bulges binned by morphological type}
\begin{center}

\begin{tabular}{c|cccc|c}
& {\bf 1} & {\bf 2}$^1$ & {\bf 3} & {\bf 4} & $\Sigma$ \\
\hline
S0/S0a & 5 &21 & 28 & 78 & 132 \\
Sa/Sab & 6 &13 & 23 & 52 & 94 \\
Sb/Sbc & 14 & 44 & 65 & 136 & 259 \\
Sc/Scd & 5 & 28 & 53 & 100 & 186 \\
Sd & 0 & 9 & 16 & 38 & 63 \\
\hline
$\Sigma$ & 30 & 115 & 185 & 404 & 734 \\
\% & 4.1 & 15.7 & 25.2 & 55.0 & 100 \\
\end{tabular}
\end{center}
\begin{center}$^1$: 10.4\,\% of all type {\bf 2} are {\bf 2+}.
\end{center}
\label{cla}
\end{table}

\begin{figure}[h]
\centering
\resizebox{\hsize}{!}{\psfig{figure=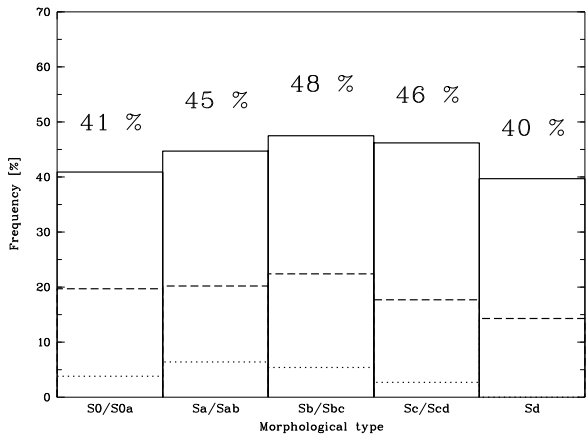,width=8.3cm,angle=270}}
\caption{
Frequency of b/p bulges binned by morphological type derived from 
Table \ref{cla}:
Dotted lines: type {\bf 1}. Dashed lines: types  {\bf 1}
+ {\bf 2}. Solid lines: all b/p types ({\bf 1} - {\bf 3}).}
\label{bpstat}
\end{figure}

734 galaxies of our sample are classifiable (Table \ref{cla}). 
From these galaxies we get a frequency of
$45.0(\pm4.5)$\,\% b/p bulges (type {\bf 1} + {\bf 2} + {\bf 3})
(Table \ref{cla}).
The distribution of all galaxies with b/p bulges binned by
morphological type shows a weak maximum at Sb/Sbc galaxies 
(Fig. \ref{bpstat}).
The smallest fraction
of b/p bulges is observed for early and late type disk
galaxies.
The fractions range from 40 to 48\,\% in the maximum.
From number statistics the
errors are 4 - 7\,\% for the frequencies of b/p bulges in each bin.
Within these errors there is no dependence of the morphological type
discernable.
Regarding only galaxies with b/p type {\bf 1} and {\bf 2} the distribution
is nearly the same.
These results are in some aspects in contrast to the former
investigations (see Sect. 5).

\subsection{CCD observations}

The optical CCD images were obtained in several observing runs
between 1985 and 1998 at Lowell Observatory (1.06m), 
Calar Alto (1.23m), and
ESO/ La Silla (0.9m, 1.54m, 2.2m, and NTT).
Standard reduction techniques for
bias subtraction and flatfielding  were
applied. Individual short exposure
frames of the galaxy  were combined. 
The data are partly published in Barteldrees \& Dettmar (\cite{bd}),
and Pohlen et al. (\cite{poh}).
Further data will be reported together
with follow-up observations of b/p bulges in
a separate paper (L\"utticke et al. in prep.).
In total we have observed 74 galaxies of our 
investigated RC3 sample up to now.

\begin{figure}[hbtp]
\centering
\resizebox{\hsize}{!}{\psfig{figure=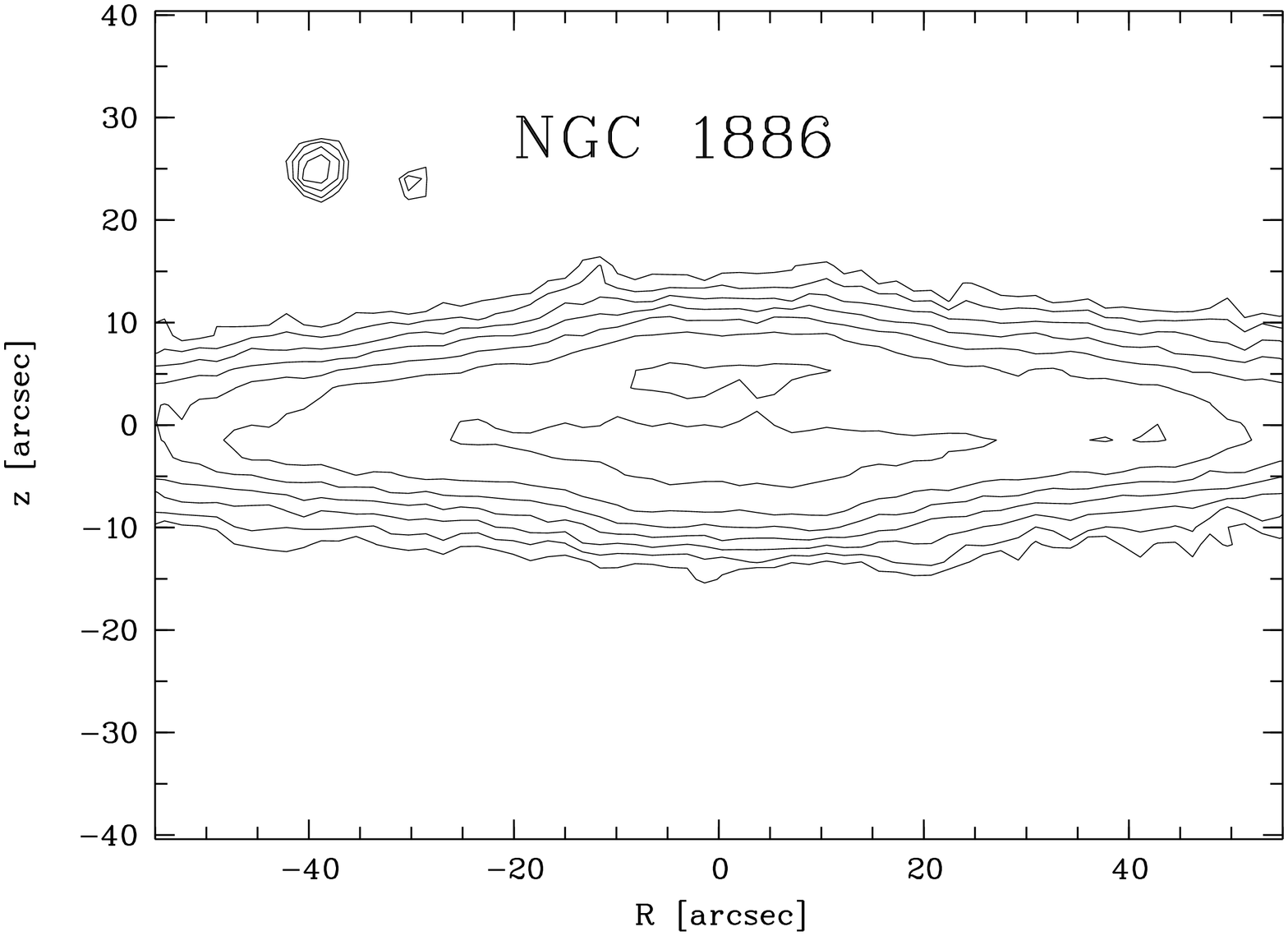,width=8.4cm,angle=270}}
\resizebox{\hsize}{!}{\psfig{figure=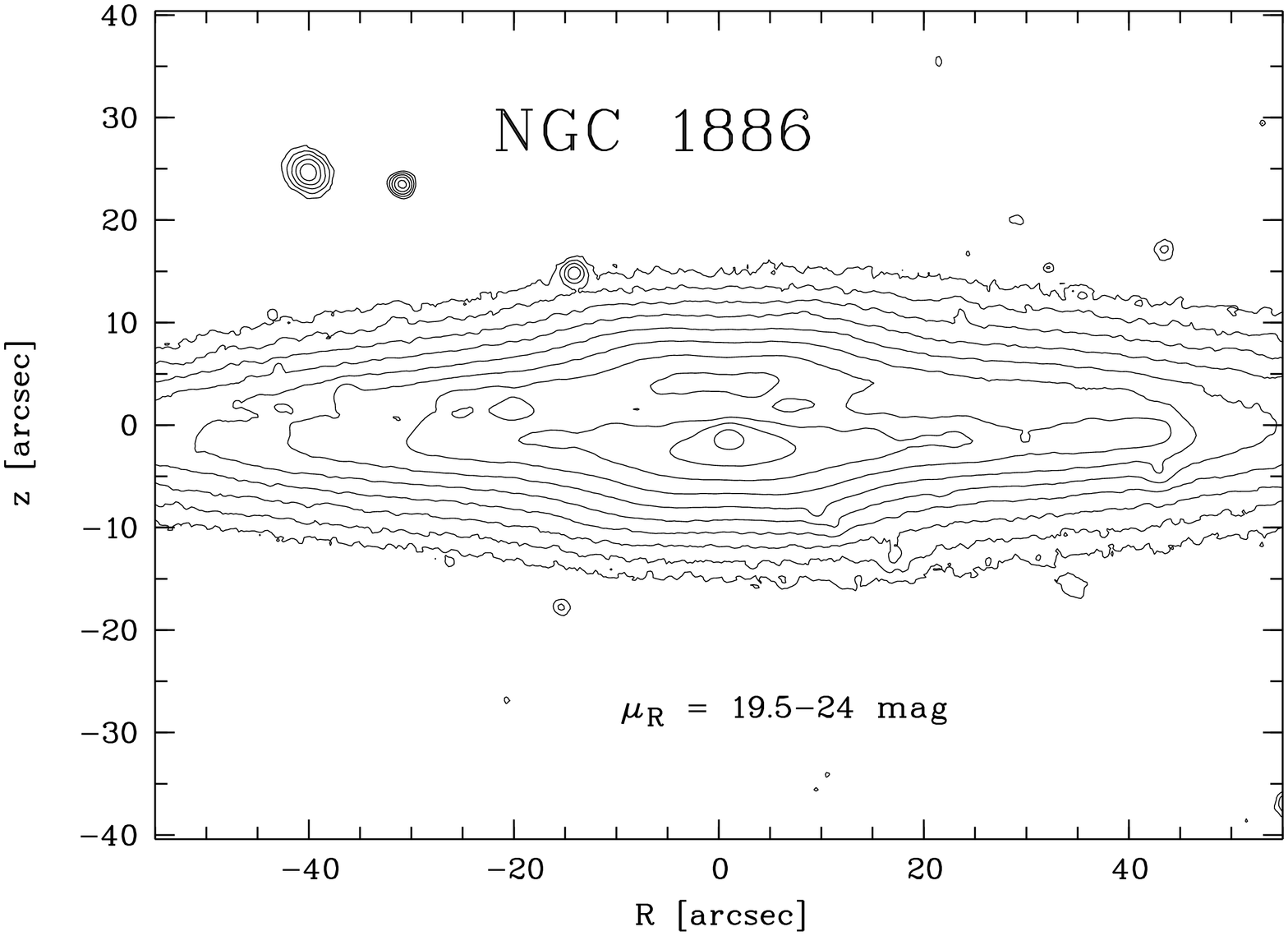,width=8.4cm,angle=270}}
\caption{Inspecting the DSS image (top)
it is  hard to decide, if there are depressions 
along the minor axis of NGC 1886 (first classified as bulge type {\bf 2+})
on both sides of the
galaxy plane.
The contour plots of the CCD images
(NGC 1886: ESO/2.2m, 30min in $r$) reveal
that NGC 1886 has a peanut bulge (bulge type {\bf 1}).}
\label{n1886}
\end{figure}

Only small differences ($<$ 10\,\%) between the classification of bulges
comparing
DSS  (RC3 sample) and CCD images
(Fig. \ref{n1886} or Fig. \ref{dss} [top] and Fig. \ref{NIR} [top]) 
are detected.  
Three bulges turned out to be type {\bf 4} rather than type {\bf 3} and
two bulges are type {\bf 3} rather than type {\bf 4}.
This could be explained by
the higher resolution of the CCD images and the unsharp nature of
the border between these
two classes. However, the total frequency of b/p bulges is not changed.
Bulges of type {\bf 2+} in the DSS images can be classified more accurately
by the high resolution of the CCD images. 
If there is any depression
along the minor axis (e.g. NGC 1886, Fig. \ref{n1886}), 
the bulge type is changed to {\bf 1}).
If not, the bulge can be classified as (type {\bf 2}).
Bulges which have only  a
depression along the minor axis
on one side are still classified as type {\bf 2+}.

The comparison of the CCD with the DSS images
shows that neither
the low resolution, nor the saturated
bright parts of some galaxies, nor the low signal-to-noise ratio, nor
unresolved structures  (small stars in the foreground)
strongly influence the classification. Therefore the results of the
statistics, which are derived from the RC3 sample and inspected with the DSS,
are strongly supported by the CCD observations.
Furthermore, the 33 galaxies 
observed in different optical filters 
reveal
the same shape of the bulge. Therefore the classification of bulges at
optical wavelengths is independent of the filter.

47 bulges of galaxies  not included in the
investigated RC3 sample (45 have $D_{25} < 2\arcmin$, additionally,
one galaxy has $\log R_{25} < 0.35$, and one S0 galaxy has $\log R_{25} < 0.30$)
are classified on CCD images of our observing runs
(Table \ref{listf}).

\subsection{NIR observations}

The essential advantage of investigations in the NIR is the small
extinction by dust at these wavelengths (Knapen et al. \cite{kna91}). Therefore
the possible influence of the dust lane near the galactic plane in edge-on
galaxies on the bulge shape is largely reduced by NIR observations.
Our NIR sample of galaxies with classifiable bulges consists of
60 galaxies (L\"utticke et al. \cite{lue2000a}, hereafter Paper II).
It reveals that 75\,\% of the bulges
have the same bulge type in the optical as well as in the NIR
(Table \ref{optNIR}). 21\,\% of the
bulges are classified in the NIR to the next lower class,
these bulges are less
box-shaped. The change of the bulge type is for two bulges in the opposite way,
one of them from a boxy to a peanut bulge likely due to the low resolution
and saturation of the DSS image of this galaxy
(NGC 5166).

\begin{table}[hbtp]
\caption{Differences in the classification of bulges in the optical and NIR}
\begin{center}
\begin{tabular}{l|cccccccccc}
opt. bulge type & 4&4&3&3&3&2&2&2&1&1 \\
NIR bulge type &4&3&4&3&2&3&2&1&2&1 \\
\hline
no. of galaxies & 12 & 1 & 7 & 9 & 0 & 5 & 12 & 1 & 0 & 9 \\
\end{tabular}
\end{center}
\label{optNIR}
\end{table}

\begin{figure}[hbtp]
\centering
\resizebox{\hsize}{!}{\psfig{figure=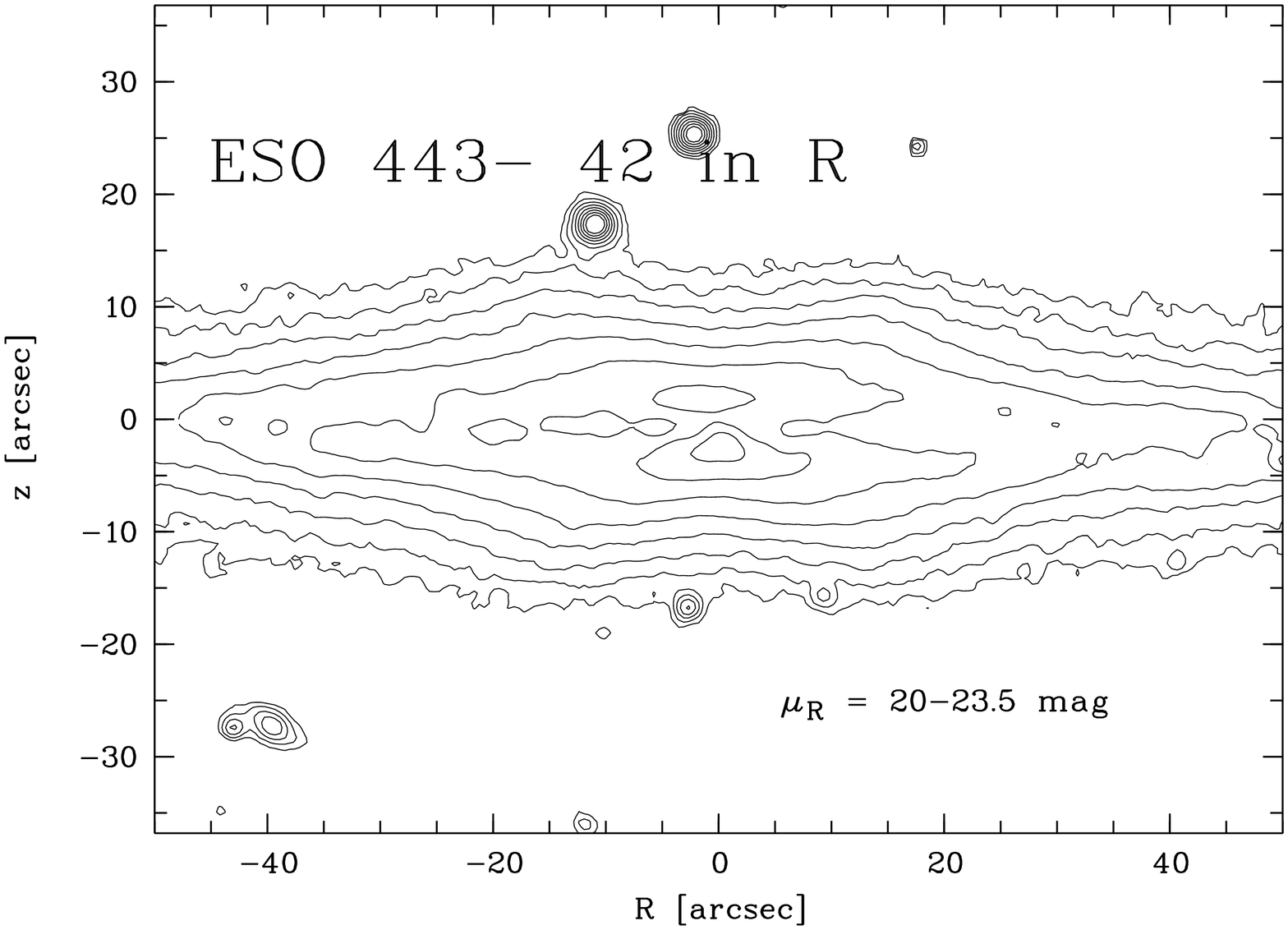,width=8.4cm,angle=270}}
\resizebox{\hsize}{!}{\psfig{figure=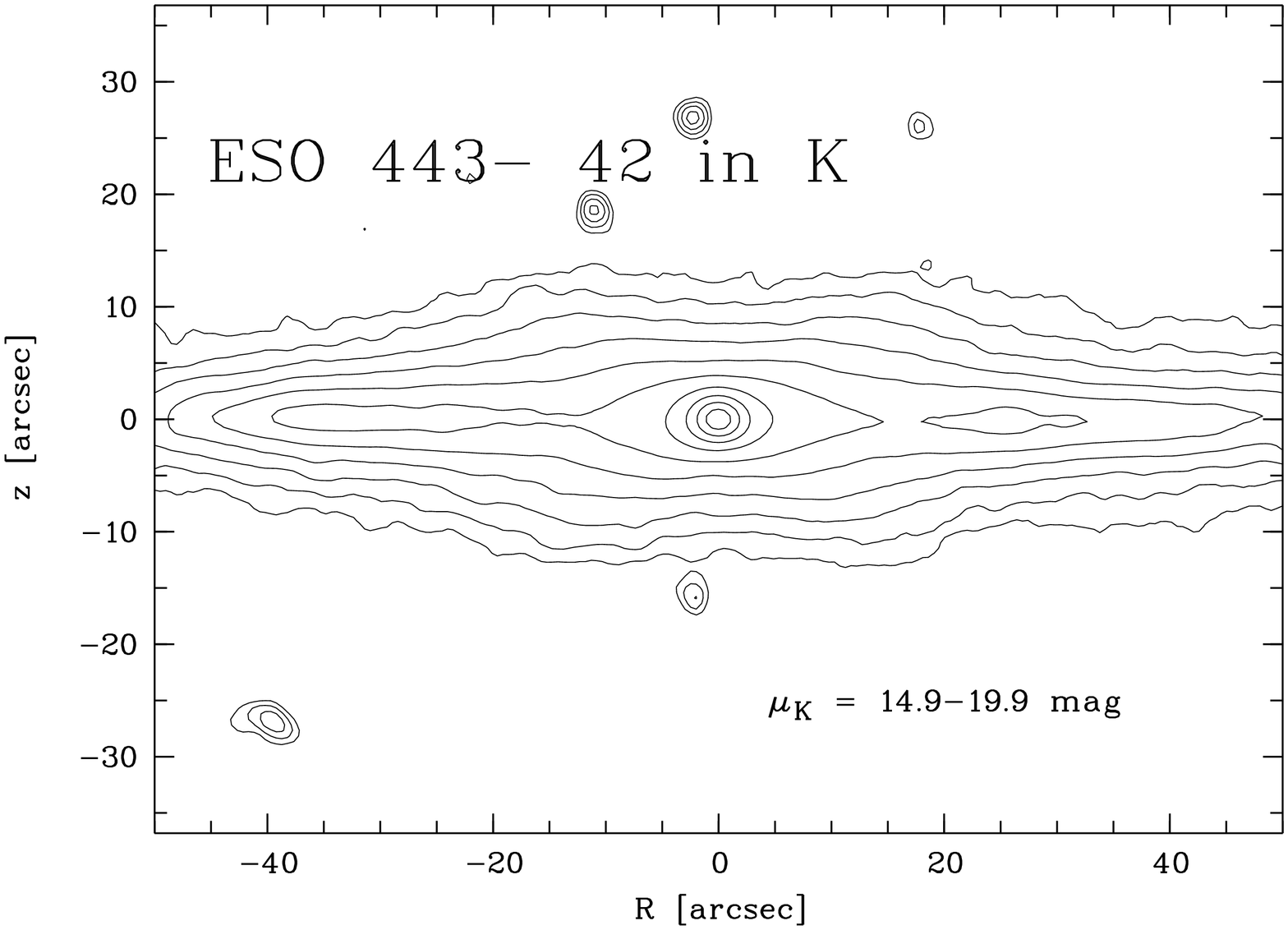,width=8.4cm,angle=270}}
\caption{The peanut bulge of ESO 443-42 is prominent in  optical CCD images
as well as in the NIR. Top: 0.9m/ESO, 10min in $R$. Bottom: 2.2m/ESO,
23min in $K'$.}
\label{NIR}
\end{figure}

All peanut bulges prominent in the optical have the same shape in
the NIR
(e.g. ESO 443-42, Fig. \ref{NIR}).
A detailed inspection
of the galaxies having less boxy bulges in the NIR shows that dust is
a possible explanation only for up to one third of these bulges.
Differences in classification can be explained mainly by general uncertainties
in the classification, the low resolution in the DSS
--- there is no difference between classifications derived
from optical CCD and NIR images ---, and low quality of some NIR images
(bad seeing).
The bulges  whose classes are changed
can frequently be marked as intermediate types (type {\bf 2}-{\bf 3} 
and {\bf 3}-{\bf 4}), respectively.
Additionally, seeing conditions around 3$\arcsec$  weaken the
boxy structure and over 50\,\%
of the bulges whose class changes were observed under such conditions.

Images of galaxies observed in all three NIR filters $J$, $H$, and $K$
reveal no difference in the bulge shape. 
Therefore
the bulge classification seems to be same in a large region of wavelengths
from the optical (350 nm) to the NIR (2.2 $\mu$m).

\section{Comparison with former statistics}

The low frequency of b/p bulges (1.2\,\%) in the sample of Jarvis (\cite{jar}) 
seems to
result from the limitation to objects of
very small diameters (Shaw \cite{sha87}). Such a restriction, coupled
with the used visual
detection method on the plates of the ESO/SERC $J$ sky survey,
results in a non-detection
of many b/p bulges and in the erroneously small
frequency of b/p bulges.
However, his division in
box ({\bf B}) and peanut-shaped ({\bf P}) bulges
is in agreement with our  
classification of b/p bulges.
Considering only galaxies included in both samples,
3/4 of the galaxies of type {\bf P} correspond to our bulge type {\bf 1}
and type {\bf B} to bulge type {\bf 2}, respectively.
All galaxies with b/p bulges listed by Jarvis (\cite{jar})
are also classified here as b/p bulges.
However, his statistics are based on a too small
number of galaxies being representative. Additionally, his sample is biased
with galaxies of known b/p bulge characteristics
 from the literature, therefore the
conclusion with regard to a dependence of b/p bulges on the Hubble type is
not meaningful.

Shaw (\cite{sha87}) uses a similar definition for his selection criterion of
b/p bulges (depression along the minor axis or a general flattening 
of the isophotes) as we do, but without classification. 
The low frequency of 20\,\%, he states, can be explained by the fact
that galaxies with b/p bulges of type {\bf 3}
are almost completely neglected.
Counting only type {\bf 1} and
{\bf 2} in our investigated
sample results also in a frequency of 20\,\% (Table \ref{cla}).
However, some prominent b/p bulges are missed
by excluding peculiar disk galaxies (e.g. NGC 3628), galaxies which are
far from edge-on (e.g. NGC 7582: Quillen et al. \cite{qui}),
or by simply ignoring others (e.g.
NGC 2424: Pohlen et al. \cite{poh}).
On the other side there are two galaxies (NGC 1596 and
NGC 4958) in his list of b/p bulges which have obviously an elliptical bulge.
The sample also shows a maximum  in the distribution of galaxies with
b/p bulges
concerning the morphological type of these galaxies
at Sb/Sbc galaxies (Table \ref{dis}). 
However, the frequencies in the individual bins
are much smaller reflecting the total low frequency of b/p bulges.
Additionally, the lack of Sd galaxies and the small frequency of 
Sc/Scd galaxies with b/p bulges are remarkable.
That is in contrast to the rather high percentage of late type
galaxies with b/p bulges which we have found in our study.
The low frequency of Sc/Scd galaxies and the lack of Sd galaxies with b/p bulges
are likely a result of
the faint nature of bulges and therefore also of their b/p structures.
Our method of detection using
the DSS and new data analysis systems could explain the differences
to Shaw (\cite{sha87}) in the bins of the late types, 
although both investigations are
based on the
same photographic material.

SA87 define three classes of galaxies marked by the degree of separation
between disk and bulge, and by the form of the bulge.
Additional to the mixture of these two criteria, it must be mentioned that
their class III is defined as ``cigar shaped'' bulge. Therefore this class
has nothing in common with b/p bulges.
This reveals
their prototype for class III (NGC 1380) and the high fraction of
elliptical bulges (68\,\%) in the class III sample (Table \ref{SA}).
Therefore the title of their list ``box-shaped galaxies'' 
is misleading. However, it should be clear by their definition of the classes
that they do not present a list and  classification of b/p  bulges.
This was and is
still  misunderstood in the literature (e.g. Bureau \& Freeman \cite{bur99}).
Therefore a strong correlation between their types
and our new introduced bulge types of the
galaxies in both samples cannot be expected
and is indeed not present (Table \ref{SA}).
However, their type I and I-II (``clear rectangular or peanut shape'')
should be classified also in our system  as
b/p bulge, but four galaxies ($\sim$\,25\,\%) out of these types
have clear elliptical bulges
(e.g. NGC 6504, Fig. \ref{dss}, bottom).
They also missed, likely due to the bad resolution of the images,
some bright galaxies with
prominent b/p bulges
(e.g. NGC 3079: Shaw et al. \cite{sha93}, Veilleux et al. \cite{vei};
NGC 7582: Quillen et al. \cite{qui}). Therefore their list is far from being
complete 
and
the frequency of b/p galaxies derived by SA87 is not comparable to
the frequency of b/p bulges derived in our investigation. 
The statistics with regard to the Hubble type are
not comparable due to the different definition of b/p bulges.
In this way it is not surprising that
the maximum in the distribution of b/p galaxies is
reached for S0 galaxies and clearly
lower values for intermediate types (Sa/Sb) in SA87 (Table \ref{dis})
are not detectable in our distribution of b/p bulges.

\begin{table}[hbtp]
\caption{Our classification compared to SA87's classification}
\begin{center}

\begin{tabular}{cr|cccc}
 & & \multicolumn{4}{c}{bulge type} \\
 & & {\bf 1} & {\bf 2} & {\bf 3} & {\bf 4} \\
\hline 
 &  I & 6 & 2 & 4 & 4 \\ 
\raisebox{1.5ex}[-1.5ex]{galaxy class} &  II & 4 & 9 & 8 & 12 \\ 
\raisebox{1.5ex}[-1.5ex]{from SA87} &  III & 0 & 2 & 4 & 13 \\ 
\end{tabular}
\end{center}
\begin{center} Listed are the numbers of galaxies. 
\end{center}
\label{SA}
\end{table}

Our derived frequency of b/p bulges  is in excellent agreement  with
the result of Dettmar \& Barteldrees (\cite{det88})
and Dettmar (\cite{det89}) who find a frequency of $45(\pm8)$\,\%.
Their detection method by use of CCD images  seems to
prevent a misclassification. All five galaxies 
in their sample of b/p bulges (published in Shaw et al. \cite{sdb}) and
our  sample have bulge types {\bf 1} or {\bf 2}.
Dettmar's (\cite{det89}, \cite{det96}) statistics concerning the Hubble type
reveal also a maximum in the distribution of galaxies with b/p bulges
at Sb/Sbc galaxies.
However, the
differences between the individual bins are much larger
than in our statistics (Table \ref{dis}). 
This can be explained by statistical errors due to the small sample size of
73 galaxies.
Dettmar (\cite{det89}, \cite{det96}) has also not found any Sd galaxy
with b/p bulge in contrast to the frequency of 40\,\% Sd
galaxies with b/p bulges in our study.
However, the lack of Sd galaxies is likely  a result of the small number of
investigated Sd galaxies in his sample.
Additionally, the already mentioned
faint nature of bulges of late type galaxies and
the uncertainties in the classification of bulges increasing
to later types (see above) lead to the fact that
the boxiness of late type bulges could be missed.
Even in small samples, where the 
possibility of comparison between the different forms of Sd bulges is not
given, the detection of b/p bulges could be influenced.
No detection of a type {\bf 1} b/p bulge in galaxies later than Sc in our 
sample could have similar reasons, if bulge type {\bf 1} in late type
galaxies exists at all.

Galaxies in previously studied samples, which do not fulfill the selection
criterion of our investigated sample due to 
morphological misclassifications of the Hubble type or
too small diameters or axis ratios,
are listed in Table \ref{listf}. However, most of these galaxies have 
$D_{25} < 2\arcmin$  and therefore their classification with the DSS
and ESO Lauberts-Valentijn Archive is uncertain.

\begin{table}[hbtp]
\caption{Comparison of different statistics: frequencies of b/p bulges
according to their morphological type}
\begin{center}
\begin{tabular}{l|ccccc}
 & S0/S0a & Sa/Sab & Sb/Sbc & Sc/Scd & Sd\\
\hline
SA87 & 33\% & 13\% & 15\% & 3\% & $-^1$ \\
Shaw$^2$   & 25\% & 33\% & 36\% & 8\% & 0\% \\
Dettmar$^3$ & 46\% & 29\% & 61\% & 42\% & 0\% \\
this study & 41\% & 45\% & 48\% & 46\% & 40\% \\
\end{tabular}
\end{center}
$^1$: no Sd galaxy in sample; 
$^2$: frequencies derived from Table 2 of Shaw \cite{sha87};
$^3$: frequencies derived from 
Figure 1 of Dettmar \cite{det89})
\label{dis}
\end{table}

\section{Discussion} 

The new classification of bulges
is very similar to the types used by 
Jarvis (\cite{jar}) and Shaw (\cite{sha87}).
The frequency of 45\,\% b/p bulges is consistent with
Dettmar \& Barteldrees (\cite{det88}), Dettmar (\cite{det89}) and
the frequency of 20\,\% prominent b/p bulges with
Shaw (\cite{sha87}).
However, the
derived frequency is now based on a much larger sample of 734 galaxies.
Furthermore, for the first time a large fraction of b/p bulges
in galaxies as late as Sd is found, and some previously unknown
peanut bulges (type 1)
are listed (Table \ref{listrc3} and \ref{listf}).

The large fraction of b/p bulges (45\,\%) 
shows that such bulges are not that peculiar
but rather quite normal. Therefore very common processes are required
to explain the origin of b/p bulges.

\begin{figure}[hbtp]
\centering
\resizebox{\hsize}{!}{\psfig{figure=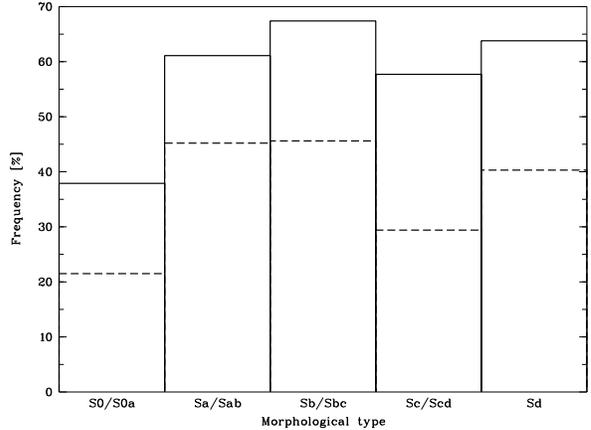,width=8.4cm,angle=270}}
\caption{Frequency of barred galaxies binned by morphological type.
 All face-on galaxies of the RC3 are included, i.e.,  all
galaxies with a diameter ($D_{25}$) larger than  1$\arcmin$  and an axis
ratio smaller than 1/3 ($\log R_{25} < 0.477$). Solid lines: strongly
barred (SB) plus
weakly barred (SX) galaxies. Dashed lines: only SB galaxies.}
\label{barstat}
\end{figure}

In order to check possible formation 
processes relating b/p bulges with bars,
we have compared the frequency
distributions of galaxies with b/p bulges and barred
galaxies.
The frequency distribution of barred
galaxies is derived from RC3 (only face-on galaxies) and contains 8587 galaxies.
(Fig. \ref{barstat}).
Both distributions binned by
morphological type show the same general dependence.
The maximum for barred galaxies is also  at Sb/Sbc
and the minimum at S0/S0a galaxies.
This minimum is more distinct for barred galaxies as
for galaxies with b/p bulge.
However, for all galaxies the fraction of barred galaxies is 55\,\%
(nearly 2/3 of them are strongly barred [SB] and the rest weakly barred
[SAB = SX])
in relation to 45\,\% of galaxies with b/p bulges. This high
percentage of barred
galaxies is also recently confirmed by Knapen et al. (\cite{kna2000}).
Furthermore, it is remarkable that
the percentage of barred Sd galaxies is relatively high
compared to
the minimal frequency of galaxies with b/p bulges for Sd galaxies.
However, it should be mentioned that our error of this bin is the largest 
due to low number statistics (Table \ref{cla}).
Other statistics of barred galaxies taken from older catalogues
(Sellwood \& Wilkinson \cite{sel}) show the same general dependences, although
the percentages vary mainly due to different fractions of weakly barred
galaxies in the catalogues. 
However, former samples
contain a much smaller number of galaxies (Sellwood \& Wilkinson \cite{sel}).
Unfortunately, the galaxies are not all classified as
SB, SX, or SA (unbarred galaxy), but there is also a
group of galaxies without any classification (S. = 28\,\%). These galaxies can
be likely interpreted as unbarred galaxies (as done in Fig. \ref{barstat}),
because in former times unbarred galaxies were simply called ``S'',
but they add uncertainties
to the quantitative conclusion.

Bars as origin for b/p bulges (Combes et al. \cite{com90},
Raha et al. \cite{rah}, Pfenniger \& Friedli \cite{pfe}) are supported by
the similarity in frequency distributions of galaxies with b/p bulges and barred
galaxies. The fraction of b/p bulges is large enough to explain the
b/p bulges by bars.
The higher frequency of barred galaxies can easily be explained by
the aspect angle of bars. End-on bars result for edge-on galaxies in
elliptical shaped bulges, bars with intermediate aspect angles in boxy
bulges, and edge-on bars in peanut bulges
(Combes et al. \cite{com90}, Pfenniger \& Friedli \cite{pfe},
Paper II).
Therefore our statistical results are consistent with recent studies
stating a strong correlation of bars and b/p bulges.
Bureau \& Freeman (\cite{bur99}) and Merrifield \& Kuij\-ken (\cite{mer})
find in observations of gas kinematics the characteristic 
``figure-of-eight'' rotation curve, which
is a strong signature for the presence of a bar, in many galaxies with
b/p bulges. 
Direct kinematic evidence for streaming motions of a bar 
in two galaxies with a b/p bulge are
reported by Veilleux et al. (\cite{vei}).
Kinematical bar diagnostics in edge-on spiral galaxies using simulations
of families
of periodic orbits and hydrodynamical simulations confirm the connection
between bars and b/p bulges
(Bureau \& Athanassoula \cite{ba}, Athanassoula \& Bureau \cite{ab}).
Additionally, we find in our NIR study 
(Paper II)
a strong correlation of bar signatures with b/p bulges. 
However, by finding a few bulges with a very complex morphology,
we do not exclude that some b/p bulges result from a recent
merger event
(Dettmar \& L\"utticke \cite{dl}, 
Paper III).

The optical
CCD images verify our classification of bulges and thereby the
statistics derived by the DSS images. 
For the analysis of faint structures and
more objective
parameters of b/p bulges (e.g. parameter for
the depression at the minor axis) CCD images are necessary.
The best parameter for
an objective classification of bulges would be
the minimum of the $a_4$ parameter determined in the NIR
(minimizing the dust influence) and after subtraction of a modelled disk and
bar. However, models of high quality need images with a high signal-to-noise 
ratio.
This would result in unreasonably observing time for the whole RC3 sample.
Therefore the classification by visual
inspection seems to be the best method until CCD data of the whole sky are
available (e.g. for the northern sky: Sloan Digital Sky Survey).

Furthermore, it is shown by NIR observations
that dust is not an important factor
for the classification of b/p bulges. Therefore they are
indeed present in disks of late type spirals as pointed out 
previously by Dettmar \& Ferrara (\cite{df}). This result is not in contrast to
Baggett \& MacKenty (\cite{bag}) who verify b/p structures in the NIR only
in four out of six galaxies because their sample contains
several misclassifications from older literature lists (section 2.1 and 5).
The comparison of optical and NIR images of b/p bulges reveals
that there is no  difference between  the shape of bulges in
different wavelengths bands
or the differences are smaller than the uncertainties
of the classification.

Additionally, as a byproduct
a new catalogue of edge-on galaxies with $D_{25} > 2\arcmin$
confirmed by visual inspection
is formed by our classification of bulges.
This method gives a homogeneous sample
of highly inclined galaxies compared to samples selected
only by axis ratio and morphological type.
This catalogue includes all galaxies with a
bulge of type 1 -- 4 and 5.2 -- 5.5 (Table \ref{listrc3}).

\section{Conclusion}
The bulge shape of a large sample of edge-on galaxies was investigated
using DSS, CCD images in the optical wavelengths regime and in the NIR. 
Our main conclusions are:
\begin{itemize}
\item We have presented a visual classification of bulges in three
b/p types and one elliptical type.

\item 45\,\% of all disk galaxies (S0 - Sd) have a b/p bulge. This fraction is
based on a sample of 734 galaxies.

\item For the first time a large fraction (40\,\%) of b/p bulges in Sd 
galaxies is found.

\item Comparisons between the fraction of barred galaxies and galaxies
with b/p bulges show that b/p bulges can be explained by
evolution scenarios based on bars.

\item NIR observations reveal that 
dust is not responsible for the shape of b/p bulges.

\item The shape of bulges is constant over a large region
of wavelengths (350 nm - 2.2 $\mu$m).

\end{itemize}

\begin{acknowledgements}
Part of this work was supported by the 
\emph{Deut\-sche For\-schungs\-ge\-mein\-schaft,
DFG\/}. This research has made use of the NASA/IPAC Extragalactic Database
(NED) which is operated by the Jet Propulsion Laboratory, California
Institute of Technology, under contract with the National Aeronautics and
Space Administration. It also uses the Digitized Sky Survey (DSS) based
on photographic data obtained using Oschin Schmidt Telescope on 
Palomar Mountain and The UK Schmidt Telescope and produced at the 
Space Telescope Science Institute. 
\end{acknowledgements}

\begin{table*}[htbp]
\begin{center}
\caption{List of galaxies in the RC3-sample with bulge classification}
\label{listrc3}
\begin{tabular}{l|cccccrcc}
\multicolumn{1}{c|}{(1)} & (2) & (3) & (4) & (5) & (6) &
\multicolumn{1}{c}{(7)} & (8) & (9)  \\
\multicolumn{1}{c|}{Object} & RA & DEC & bulge & NIR & earlier &
\multicolumn{1}{c}{$T$} & log & log  \\
 & (2000) & (2000) & type & bulge & detections & & $D_{25}$ & $R_{25}$  \\
 & \,  & \, & \, & type & & \, & \, \\
\hline 
IC   5376 & 00 01 & +34 31 & 4 &    & & 2.0 & 1.30 & 0.76\\
NGC  7814 & 00 03 & +16 08 & 4$^f$ &    & & 2.0 & 1.74 & 0.38\\
NGC  7817 & 00 03 & +20 45 & 5.1 &    &  & 4.0 & 1.55 & 0.58\\
ESO  293- 34     &       00          06 &$-$41          29 &     5.4&&&  6.1 &
1.50& 0.51\\
\end{tabular}
\end{center}
\end{table*}

\begin{table*}[hbtp]
Table \ref{listrc3} continued.
\begin{center}
\begin{tabular}{l|cccccrcc}
\multicolumn{1}{c|}{(1)} & (2) & (3) & (4) & (5) & (6) &
\multicolumn{1}{c}{(7)} & (8) & (9)  \\
\multicolumn{1}{c|}{Object} & RA & DEC & bulge & NIR & earlier &
\multicolumn{1}{c}{$T$} & log & log  \\
 & (2000) & (2000) & type & bulge & detections & & $D_{25}$ & $R_{25}$  \\
 & \,  & \, & \, & type & & \, & \, \\
\hline 
NGC   493          &      01          22&  +00          56 &     3 &&& 6.0 &1.53
& 0.51\\
NGC   497          &      01          22& $-$00          52 &     5.1 &&& 4.0 &
1.32& 0.38\\
NGC   494          &      01          22&  +33          10 &     5.1 &&& 2.0 &
1.30& 0.41\\
NGC   517          &      01          24&  +33          25 &     5.1 &&&-2.0 &
1.30& 0.30\\
\end{tabular}
\end{center}
\end{table*}

\begin{table*}[hbtp]
Table \ref{listrc3} continued.
\begin{center}
\begin{tabular}{l|cccccrcc}
\multicolumn{1}{c|}{(1)} & (2) & (3) & (4) & (5) & (6) &
\multicolumn{1}{c}{(7)} & (8) & (9)  \\
\multicolumn{1}{c|}{Object} & RA & DEC & bulge & NIR & earlier &
\multicolumn{1}{c}{$T$} & log & log  \\
 & (2000) & (2000) & type & bulge & detections & & $D_{25}$ & $R_{25}$  \\
 & \,  & \, & \, & type & & \, & \, \\
\hline
NGC   908          &      02          23& $-$21          14 &     5.1 &&& 5.0 &1.78& 0.36\\
              UGC  1856         &      02          24&  +31          36 &     4 &&& 7.0 &1.37& 0.97\\
              UGC  1883         &      02          25&  +11          28 &     4 &&& 7.0 &1.34& 0.58\\
NGC   931          &      02          28&  +31          18 &     4 &&& 4.0 &1.59& 0.67\\
\end{tabular}
\end{center}
\end{table*}

\begin{table*}[hbtp]
Table \ref{listrc3} continued.
\begin{center}
\begin{tabular}{l|cccccrcc}
\multicolumn{1}{c|}{(1)} & (2) & (3) & (4) & (5) & (6) &
\multicolumn{1}{c}{(7)} & (8) & (9)  \\
\multicolumn{1}{c|}{Object} & RA & DEC & bulge & NIR & earlier &
\multicolumn{1}{c}{$T$} & log & log  \\
 & (2000) & (2000) & type & bulge & detections & & $D_{25}$ & $R_{25}$  \\
 & \,  & \, & \, & type & & \, & \, \\
\hline

NGC  1301           &     03          20& $-$18          42&      2 &&& 3.5 &1.34 &0.69\\
              UGC  2709          &     03          23& +38          40&      4 &&& 3.0 &1.37 &0.48\\
NGC  1325           &     03          24& $-$21          32&      5.1 &&& 4.0 &1.67 &0.47\\
NGC  1324           &     03          25& $-$05          44&      3 &&& 3.0 &1.32 &0.42\\
\end{tabular}
\end{center}
\end{table*}

\begin{table*}[hbtp]
Table \ref{listrc3} continued.
\begin{center}
\begin{tabular}{l|cccccrcc}
\multicolumn{1}{c|}{(1)} & (2) & (3) & (4) & (5) & (6) &
\multicolumn{1}{c}{(7)} & (8) & (9)  \\
\multicolumn{1}{c|}{Object} & RA & DEC & bulge & NIR & earlier &
\multicolumn{1}{c}{$T$} & log & log  \\
 & (2000) & (2000) & type & bulge & detections & & $D_{25}$ & $R_{25}$  \\
 & \,  & \, & \, & type & & \, & \, \\
\hline

 UGC  3042          &     04          28& +70          20&      5.1  &&&4.0 &1.33 &0.45\\
NGC  1589           &     04          30& +00          51&      2$^c$  &&SA III&2.2 &1.50 &0.49\\
IC   2085           &     04          31& $-$54          25&      4 &&&-2.0 &1.38 &0.67\\
              ESO  202- 35       &     04          32& $-$49          40&      3  &&&3.0 &1.42 &0.86\\
\end{tabular}
\end{center}
\end{table*}

\begin{table*}[hbtp]
Table \ref{listrc3} continued.
\begin{center}
\begin{tabular}{l|cccccrcc}
\multicolumn{1}{c|}{(1)} & (2) & (3) & (4) & (5) & (6) &
\multicolumn{1}{c}{(7)} & (8) & (9)  \\
\multicolumn{1}{c|}{Object} & RA & DEC & bulge & NIR & earlier &
\multicolumn{1}{c}{$T$} & log & log  \\
 & (2000) & (2000) & type & bulge & detections & & $D_{25}$ & $R_{25}$  \\
 & \,  & \, & \, & type & & \, & \, \\
\hline
              UGC  3403           &    06          10& +71          23   &  5.2 &&& 6.0 &1.36 &0.52\\
              UGC  3425           &    06          14& +66          34   &  4$^f$ &&& 3.0 &1.40 &0.86\\
              UGC  3433           &    06          15& +00          11   &  4 &&& 4.0 &1.31 &0.51\\
              UGC  3420           &    06          16& +75          56   &  4 &&& 3.0 &1.41 &0.49\\
\end{tabular}
\end{center}
\end{table*}

\clearpage

\begin{table*}[hbtp]
Table \ref{listrc3} continued.
\begin{center}
\begin{tabular}{l|cccccrcc}
\multicolumn{1}{c|}{(1)} & (2) & (3) & (4) & (5) & (6) &
\multicolumn{1}{c}{(7)} & (8) & (9)  \\
\multicolumn{1}{c|}{Object} & RA & DEC & bulge & NIR & earlier &
\multicolumn{1}{c}{$T$} & log & log  \\
 & (2000) & (2000) & type & bulge & detections & & $D_{25}$ & $R_{25}$  \\
 & \,  & \, & \, & type & & \, & \, \\
\hline

              ESO   89- 12    &       08          10&$-$64          56   &  4&&& 4.0& 1.36& 0.79\\
              UGC  4257       &       08          10&+24          53   &  5.5&&& 6.0& 1.33& 1.05\\
              ESO  561- 23    &       08          11&$-$18          18   &  4&&& 3.7& 1.31& 0.81\\
NGC  2523B      &       08          12&+73          33   &  4&&& 3.0& 1.33& 0.83\\
\end{tabular}
\end{center}
\end{table*}

\begin{table*}[hbtp]
Table \ref{listrc3} continued.
\begin{center}
\begin{tabular}{l|cccccrcc}
\multicolumn{1}{c|}{(1)} & (2) & (3) & (4) & (5) & (6) &
\multicolumn{1}{c}{(7)} & (8) & (9)  \\
\multicolumn{1}{c|}{Object} & RA & DEC & bulge & NIR & earlier &
\multicolumn{1}{c}{$T$} & log & log  \\
 & (2000) & (2000) & type & bulge & detections & & $D_{25}$ & $R_{25}$  \\
 & \,  & \, & \, & type & & \, & \, \\
\hline
NGC  2798          &     09          17& +42          00  &    5.1&&&  1.0& 1.41& 0.42\\
UGC  4906          &     09          17& +52          59  &    4&&SA III&  1.0& 1.30& 0.59\\
              ESO  433- 12       &     09          18& $-$32          28  &    5.1&&&  3.7& 1.36& 0.69\\
IC   2461          &     09          19& +37          11  &    4&&&  3.0& 1.37& 0.75\\
\end{tabular}
\end{center}
\end{table*}

\begin{table*}[hbtp]
Table \ref{listrc3} continued.
\begin{center}
\begin{tabular}{l|cccccrcc}
\multicolumn{1}{c|}{(1)} & (2) & (3) & (4) & (5) & (6) &
\multicolumn{1}{c}{(7)} & (8) & (9)  \\
\multicolumn{1}{c|}{Object} & RA & DEC & bulge & NIR & earlier &
\multicolumn{1}{c}{$T$} & log & log  \\
 & (2000) & (2000) & type & bulge & detections & & $D_{25}$ & $R_{25}$  \\
 & \,  & \, & \, & type & & \, & \, \\
\hline

              ESO  263- 15     &       10          12&$-$47          17&     4 &&& 6.0 &1.46 &0.94\\
              ESO  436-  1     &       10          12&$-$27          50&     5.2 &&& 4.3 &1.46 &0.92\\
NGC  3153         &       10          12&+12          40&     5.1 &&& 6.0 &1.33 &0.36\\
              ESO  263- 18     &       10          13&$-$43          43&     4 &&& 4.0 &1.31 &0.77\\
\end{tabular}
\end{center}
\end{table*}

\begin{table*}[hbtp]
Table \ref{listrc3} continued.
\begin{center}
\begin{tabular}{l|cccccrcc}
\multicolumn{1}{c|}{(1)} & (2) & (3) & (4) & (5) & (6) &
\multicolumn{1}{c}{(7)} & (8) & (9)  \\
\multicolumn{1}{c|}{Object} & RA & DEC & bulge & NIR & earlier &
\multicolumn{1}{c}{$T$} & log & log  \\
 & (2000) & (2000) & type & bulge & detections & & $D_{25}$ & $R_{25}$  \\
 & \,  & \, & \, & type & & \, & \, \\
\hline
NGC  3413          &       10          51&+32          46&     4 &&&-2.0 &1.34 &0.38\\
              ESO  569- 14     &       10          51&$-$19          53&     3 &&& 6.2 &1.52 &0.85\\
              UGC  5974        &       10          51&+04          34&     5.1 &&& 6.0 &1.31 &0.40\\
PGC 32573                     &       10          51&$-$17          07&     5.1 &&& 1.7 &1.34 &0.39\\
\end{tabular}
\end{center}
\end{table*}

\begin{table*}[hbtp]
Table \ref{listrc3} continued.
\begin{center}
\begin{tabular}{l|cccccrcc}
\multicolumn{1}{c|}{(1)} & (2) & (3) & (4) & (5) & (6) &
\multicolumn{1}{c}{(7)} & (8) & (9)  \\
\multicolumn{1}{c|}{Object} & RA & DEC & bulge & NIR & earlier &
\multicolumn{1}{c}{$T$} & log & log  \\
 & (2000) & (2000) & type & bulge & detections & & $D_{25}$ & $R_{25}$  \\
 & \,  & \, & \, & type & & \, & \, \\
\hline

NGC  3717         &      11          31& $-$30          18&      2+  &&& 3.0 &1.78& 0.73\\
UGC  6534         &      11          33& +63          16&      5.1  &&& 7.0 &1.40& 0.61\\
NGC  3749         &      11          35& $-$37          59&      3  &&& 1.0 &1.50& 0.60\\
NGC  3735         &      11          35& +70          32&      3  &&& 5.0 &1.62& 0.70\\
\end{tabular}
\end{center}
\end{table*}

\clearpage

\begin{table*}[hbtp]
Table \ref{listrc3} continued.
\begin{center}
\begin{tabular}{l|cccccrcc}
\multicolumn{1}{c|}{(1)} & (2) & (3) & (4) & (5) & (6) &
\multicolumn{1}{c}{(7)} & (8) & (9)  \\
\multicolumn{1}{c|}{Object} & RA & DEC & bulge & NIR & earlier &
\multicolumn{1}{c}{$T$} & log & log  \\
 & (2000) & (2000) & type & bulge & detections & & $D_{25}$ & $R_{25}$  \\
 & \,  & \, & \, & type & & \, & \, \\
\hline

NGC  4094     &          12          05 &$-$14          31   &   5.1 &&&  5.5 &1.62 &0.45\\
              UGC  7086     &          12          05 &+77          30   &   4 &&&  3.0 &1.43 &0.87\\
NGC  4096     &          12          06 &+47          28   &   5.4 &&&  5.0 &1.82 &0.57\\
NGC  4100     &          12          06 &+49          34   &   5.1 &&&  4.0 &1.73 &0.48\\
\end{tabular}
\end{center}
\end{table*}

\begin{table*}[hbtp]
Table \ref{listrc3} continued.
\begin{center}
\begin{tabular}{l|cccccrcc}
\multicolumn{1}{c|}{(1)} & (2) & (3) & (4) & (5) & (6) &
\multicolumn{1}{c}{(7)} & (8) & (9)  \\
\multicolumn{1}{c|}{Object} & RA & DEC & bulge & NIR & earlier &
\multicolumn{1}{c}{$T$} & log & log  \\
 & (2000) & (2000) & type & bulge & detections & & $D_{25}$ & $R_{25}$  \\
 & \,  & \, & \, & type & & \, & \, \\
\hline

NGC  4338      &         12          23 &+28          53  &    4 &&&  7.0 &1.34 &1.01\\
NGC  4346      &         12          23 &+46          59  &    4 &&SA III& -2.0 &1.52 &0.42\\
NGC  4343      &         12          23 &+06          57  &    4 &&&  3.0 &1.40 &0.54\\
IC   3253      &         12          23 &$-$34          37  &    5.1 &&&  4.8 &1.45 &0.39\\
\end{tabular}
\end{center}
\end{table*}

\clearpage

\begin{table*}[hbtp]
Table \ref{listrc3} continued.
\begin{center}
\begin{tabular}{l|cccccrcc}
\multicolumn{1}{c|}{(1)} & (2) & (3) & (4) & (5) & (6) &
\multicolumn{1}{c}{(7)} & (8) & (9)  \\
\multicolumn{1}{c|}{Object} & RA & DEC & bulge & NIR & earlier &
\multicolumn{1}{c}{$T$} & log & log  \\
 & (2000) & (2000) & type & bulge & detections & & $D_{25}$ & $R_{25}$  \\
 & \,  & \, & \, & type & & \, & \, \\
\hline

NGC  4565      &         12          36 &+25          59  &    2$^a$ &&J P, SA II, S&  3.0 &2.20 &0.87\\
              UGC  7774      &         12          36 &+40          00  &    5.4 &&&  7.0 &1.56 &0.92\\
NGC  4568      &         12          36 &+11          14  &    5.1 &&&  4.0 &1.66 &0.36\\
NGC  4570      &         12          36 &+07          14  &    4 &&& -2.0 &1.58 &0.52\\
\end{tabular}
\end{center}
\end{table*}

\begin{table*}[hbtp]
Table \ref{listrc3} continued.
\begin{center}
\begin{tabular}{l|cccccrcc}
\multicolumn{1}{c|}{(1)} & (2) & (3) & (4) & (5) & (6) &
\multicolumn{1}{c}{(7)} & (8) & (9)  \\
\multicolumn{1}{c|}{Object} & RA & DEC & bulge & NIR & earlier &
\multicolumn{1}{c}{$T$} & log & log  \\
 & (2000) & (2000) & type & bulge & detections & & $D_{25}$ & $R_{25}$  \\
 & \,  & \, & \, & type & & \, & \, \\
\hline

              UGC  8067      &         12          57 &$-$01          42  &    4 &&&  3.0 &1.32 &0.73\\
NGC  4835A     &         12          57 &$-$46          22  &    2$^d$ &&SDB&  6.0 &1.41 &0.80\\
              MCG -1-33- 60  &         12          57 &$-$09          38  &    4 &&&  6.7 &1.49 &0.98\\
NGC  4843      &         12          58 &$-$03          37  &    2 &&&  0.0 &1.32 &0.64\\
\end{tabular}
\end{center}
\end{table*}

\begin{table*}[hbtp]
Table \ref{listrc3} continued.
\begin{center}
\begin{tabular}{l|cccccrcc}
\multicolumn{1}{c|}{(1)} & (2) & (3) & (4) & (5) & (6) &
\multicolumn{1}{c}{(7)} & (8) & (9)  \\
\multicolumn{1}{c|}{Object} & RA & DEC & bulge & NIR & earlier &
\multicolumn{1}{c}{$T$} & log & log  \\
 & (2000) & (2000) & type & bulge & detections & & $D_{25}$ & $R_{25}$  \\
 & \,  & \, & \, & type & & \, & \, \\
\hline

NGC  5170      &          13          29 &$-$17          57   &   3$^e$ &&&  5.0& 1.92& 0.91\\
              UGC  8498     &          13          30 &+31          37   &   4 &&&  3.0& 1.44& 0.46\\
NGC  5188      &          13          31 &$-$34          47   &   5.1 &&&  2.6& 1.48& 0.43\\
              ESO   21-  4  &          13          32 &$-$77          50   &   3$^d$ &&& -0.4& 1.35& 0.35\\
\end{tabular}
\end{center}
\end{table*}

\begin{table*}[hbtp]
Table \ref{listrc3} continued.
\begin{center}
\begin{tabular}{l|cccccrcc}
\multicolumn{1}{c|}{(1)} & (2) & (3) & (4) & (5) & (6) &
\multicolumn{1}{c}{(7)} & (8) & (9)  \\
\multicolumn{1}{c|}{Object} & RA & DEC & bulge & NIR & earlier &
\multicolumn{1}{c}{$T$} & log & log  \\
 & (2000) & (2000) & type & bulge & detections & & $D_{25}$ & $R_{25}$  \\
 & \,  & \, & \, & type & & \, & \, \\
\hline

              UGC  9182       &        14          20 &+21          56   &   3 &&&  7.0& 1.39&0.63\\
NGC  5577       &        14          21 &+03          26   &   4 &&&  4.0& 1.53&0.53\\
IC   4402       &        14          21 &$-$46          17   &   5.2 &&&  3.0& 1.62&0.69\\
NGC  5587       &        14          22 &+13          55   &   5.1 &&&  0.0& 1.42&0.50\\
\end{tabular}
\end{center}
\end{table*}

\clearpage

\begin{table*}[hbtp]
Table \ref{listrc3} continued.
\begin{center}
\begin{tabular}{l|cccccrcc}
\multicolumn{1}{c|}{(1)} & (2) & (3) & (4) & (5) & (6) &
\multicolumn{1}{c}{(7)} & (8) & (9)  \\
\multicolumn{1}{c|}{Object} & RA & DEC & bulge & NIR & earlier &
\multicolumn{1}{c}{$T$} & log & log  \\
 & (2000) & (2000) & type & bulge & detections & & $D_{25}$ & $R_{25}$  \\
 & \,  & \, & \, & type & & \, & \, \\
\hline

NGC  5878     &          15          13 &$-$14          16   &   5.1 &&&  3.0& 1.54 &0.38\\
              UGC  9780     &          15          14 &+44          35   &   2 &&&  6.0& 1.32 &0.91\\
              ESO  274-  1  &          15          14 &$-$46          48   &   5.2 &&&  6.7& 2.04 &0.84\\
NGC  5899     &          15          15 &+42           2   &   5.1 &&&  5.0& 1.50 &0.43\\
\end{tabular}
\end{center}
\end{table*}

\begin{table*}[hbtp]
Table \ref{listrc3} continued.
\begin{center}
\begin{tabular}{l|cccccrcc}
\multicolumn{1}{c|}{(1)} & (2) & (3) & (4) & (5) & (6) &
\multicolumn{1}{c}{(7)} & (8) & (9)  \\
\multicolumn{1}{c|}{Object} & RA & DEC & bulge & NIR & earlier &
\multicolumn{1}{c}{$T$} & log & log  \\
 & (2000) & (2000) & type & bulge & detections & & $D_{25}$ & $R_{25}$  \\
 & \,  & \, & \, & type & & \, & \, \\
\hline
IC   4635       &        17          15 &$-$77          29   &   2 &&&  3.3 &1.48 &0.64\\
NGC  6361       &        17          18 &+60          36   &   3 &4&&  3.0 &1.34 &0.54\\
              ESO  138- 24    &        17          24 &$-$59          22   &   5.2 &&&  5.0 &1.38 &0.63\\
NGC  6395       &        17          26 &+71          05   &   5.1 &&&  6.0 &1.38 &0.51\\
\end{tabular}
\end{center}
\end{table*}

\begin{table*}[hbtp]
Table \ref{listrc3} continued.
\begin{center}
\begin{tabular}{l|cccccrcc}
\multicolumn{1}{c|}{(1)} & (2) & (3) & (4) & (5) & (6) &
\multicolumn{1}{c}{(7)} & (8) & (9)  \\
\multicolumn{1}{c|}{Object} & RA & DEC & bulge & NIR & earlier &
\multicolumn{1}{c}{$T$} & log & log  \\
 & (2000) & (2000) & type & bulge & detections & & $D_{25}$ & $R_{25}$  \\
 & \,  & \, & \, & type & & \, & \, \\
\hline

              ESO  142- 19   &         19          33 &$-$58          06   &   4$^d$ &&&  3.0& 1.61& 0.78\\
IC   4872      &         19          35 &$-$57          31   &   3$^d$ &&&  6.7& 1.54& 0.98\\
              ESO  460- 18   &         19          38 &$-$29          48   &   5.1 &&&  4.0& 1.31& 0.46\\
              MCG -3-50-  3  &         19          42 &$-$15          52   &   5.2 &&&  5.0& 1.31& 0.79\\
\end{tabular}
\end{center}
\end{table*}

\clearpage

\begin{table*}[hbtp]
Table \ref{listrc3} continued.
\begin{center}
\begin{tabular}{l|cccccrcc}
\multicolumn{1}{c|}{(1)} & (2) & (3) & (4) & (5) & (6) &
\multicolumn{1}{c}{(7)} & (8) & (9)  \\
\multicolumn{1}{c|}{Object} & RA & DEC & bulge & NIR & earlier &
\multicolumn{1}{c}{$T$} & log & log  \\
 & (2000) & (2000) & type & bulge & detections & & $D_{25}$ & $R_{25}$  \\
 & \,  & \, & \, & type & & \, & \, \\
\hline

NGC  7064          &     21          29 &$-$52          45   &   5.4$^d$ &&&  5.0 &1.58& 0.79\\
              ESO   48-  2       &     21          36 &$-$76          20   &   2 &&&  3.0 &1.37& 0.60\\
NGC  7090          &     21          36 &$-$54          33   &   3 &&&  5.0 &1.87& 0.77\\
              MCG  2-55-  7      &     21          40 &+12          21   &   4 &&&  3 $^6$ &1.37& 0.63\\
\end{tabular}
\end{center}
\end{table*}

\begin{table*}[hbtp]
Table \ref{listrc3} continued.
\begin{center}
\begin{tabular}{l|cccccrcc}
\multicolumn{1}{c|}{(1)} & (2) & (3) & (4) & (5) & (6) &
\multicolumn{1}{c}{(7)} & (8) & (9)  \\
\multicolumn{1}{c|}{Object} & RA & DEC & bulge & NIR & earlier &
\multicolumn{1}{c}{$T$} & log & log  \\
 & (2000) & (2000) & type & bulge & detections & & $D_{25}$ & $R_{25}$  \\
 & \,  & \, & \, & type & & \, & \, \\
\hline

NGC  7416       &        22          55 &$-$05          29   &   5.1 & & &  3.0&
 1.50& 0.55\\
IC   5267A      &        22          55 &$-$43          26   &   5.1 & & &  4.3& 1.38& 0.48\\
IC   5269B      &        22          56 &$-$36          14   &   4 &&&  5.5& 1.61& 0.72\\
IC   5264       &        22          56 &$-$36          33   &   4 &&&  2.3& 1.39& 0.69\\
\end{tabular}
\end{center}
\end{table*}

\begin{table*}[hbtp]
Notes to the table: \\
\\
Col. (1): Principal name of the galaxy using
following order: NGC, IC, ESO, UGC, MCG, UGCA, and PGC. \\
\noindent  *: NGC 2788A = ESO 60- 24 (NGC number is not in RC3). \\
\noindent  **: UGC 10610 = MCG 7-35- 4 (UGC number is not in RC3).\\
\\
Col. (2): RA in hours and minutes (leaving the seconds). \\
\\
Col. (3): DEC in degree and minutes (leaving the seconds). \\
\\
Col. (4): Bulge type as defined in section 3. Numbers marked with a letter
 are confirmed by CCD observations. \\
$^a$: Calar Alto, 1.2m, 1996 \\
$^b$: ESO/La Silla, 0.9m, 1998\\
$^c$: ESO/La Silla, 1.54m, 1996, 1998 \\
$^d$: ESO/La Silla, 2.2m, 1985, 1987, 1990 \\
$^e$: ESO/La Silla, NTT, 1991 \\
$^f$: Lowell, 1.06m, 1989, 1990 \\
$^g$: ESO/ST-ECF Archive, VLT, 1999  \\
\\
Col. (5): Bulge type in the NIR 
(for details of the observations cf. Paper II).
\\
Col. (6): Earlier detections and classifications:  \\
{\bf J P}: Jarvis (\cite{jar}), peanut-shaped \\
{\bf J B}: Jarvis (\cite{jar}), box-shaped \\
{\bf SA I}: SA87 (\cite{sa87}), class I \\
{\bf SA II}: SA87 (\cite{sa87}), class II \\
{\bf SA III}: SA87 (\cite{sa87}), class III\\
{\bf S}: Shaw (\cite{sha87}) \\
{\bf SDB}: Shaw et al. (\cite{sdb}) (only literature sample)\\
{\bf Sh}: Shaw (\cite{sha93a}) (only galaxies not listed in S)\\
\\
Col. (7): Morphological type from RC3, except for \\
 $^1$: Skiff (\cite{ski}) \\
 $^2$: Own classification, only bulges of type 5
 (checked by 12 galaxies classified by Skiff (\cite{ski}),
$T_{skiff} - T_{own} \le \left| 1\right| $).
\\
 $^3$: NGC\,3079: Morphological type (SBS5) differs from $T$ (7.0) (both RC3).
\\ 
 $^4$: UGC\,8032: Morphological type from
Skiff (\cite{ski}) and Haynes et al. (\cite{hay}). The $T$ value of
Haynes et al. is based on
Kraan-Korteweg (\cite{kra}) 
who refers to a priv. communication with Sandage (1980).
Yasuda et al. (\cite{yas95}, \cite{yas97}) 
gave a wrong classification  $T\!=\!5$.
In 1995 they refer to VCC (Binggeli et al. \cite{vcc}) and in 1997 to
RC3, but in both catalogues the type of UGC\,8032 is not listed. \\
 $^5$:
``The Surface Photometry Catalogue of the ESO-Uppsala Galaxies''
(Lauberts \& Valentijn \cite{lv}). \\
 $^6$:  NASA Extragalactic Database (NED).\\
 $^7$: ``The ESO/Uppsala Survey of the ESO ($B$) Atlas'' (Lauberts \cite{lau}).
 \\
\\
Col. (8): Decimal logarithm of the 25 ($B$) isophotal major
diameter from the RC3.  \\
\\
Col. (9): Decimal logarithm of the 25 ($B$) isophotal
axis ratio ($a/b$) from the RC3.\\

\end{table*}

\begin{table*}[htbp]
\begin{center}
\caption{List of galaxies with bulge classification not included in our
selected RC3-sample}
\label{listf}
\begin{tabular}{l|cccccrll}
\multicolumn{1}{c|}{(1)} & (2) & (3) & (4) & (5) & (6) &
\multicolumn{1}{c}{(7)} & \multicolumn{1}{c}{(8)} & \multicolumn{1}{c}{(9)}\\
\multicolumn{1}{c|}{Object} & RA & DEC & bulge & NIR & earlier & \multicolumn{1}{c}{$T$} &
\multicolumn{1}{c}{log} & \multicolumn{1}{c}{log}  \\
 & (2000) & (2000) & type & bulge & detections & & \multicolumn{1}{c}{$D_{25}$}
 & \multicolumn{1}{c}{$R_{25}$}  \\
 & \,  & \, & \, & type & & \, & \, \\
\hline 

              ESO  112-  4     &       00          28 &$-$58          06  &    2$^d$  &&SDB& 5.6 &1.12& 0.85\\
              ESO  150- 14     &       00          36 &$-$56          54  &    4$^d$  &&&-0.7 &1.28& 0.99\\
ESO 195-  5 & 00 50 & $-$52 07 & 4 && J B & 1 $^6$ & 1.04$^6$ & 0.74$^6$ \\
              ESO  151-  4     &       00          56 &$-$53          11  &    1  &&J P, SDB &-2.0 &1.13& 0.56\\
              ESO  113-  6     &       01          01 &$-$59          44  &    3$^d$  &&& 5.0 &1.22& 0.97\\
NGC   489 &01  21& +09 12 & 4 &&SA II& -2 $^6$ & 1.22 & 0.65\\
              ESO  244- 48     &       01          39 &$-$47          07  &    4$^d$  &&& 3.0 &1.14& 0.82\\
              ESO  244- 49     &       01          39 &$-$46          59  &    3  &&J P&-0.6 &1.21& 0.94\\
NGC  1030        &       02          39 &+18          01  &    3$^f$  &&&-1 $^1$ &1.20& 0.38\\
NGC  1175        &       03          04 &+42          20  &    2+  &2+&J B&-1.0 &1.29& 0.50\\
              ESO  205-  9     &       05          57 &$-$52          22  &    4  &&J B& 3.7 &1.27& 0.86\\
NGC  2191        &       06          08 &$-$52          30  &    5.1  &&&-2.0 &1.24& 0.31\\
NGC  2769        &       09          10 &+50          25  &    2$^f$  &&& 1.0 &1.24& 0.62\\
ESO 433- 19 & 09 24 & $-$28 10 & 1 &&J B, SDB& 3 $^6$ & 1.23$^6$ & 0.75$^6$ \\
IC   2560        &       10          16 &$-$33          33  &  2$^d$ &&& 3.3 &1.50& 0.20\\
              ESO  376-  9     &       10          42 &$-$33          14  &    4$^d$  &&&-2.0 &1.19& 0.56\\
              ESO  318-  2     &       10          42 &$-$40          34  &    4$^d$  &&& 1.0 &1.14& 0.61\\
              ESO  376- 23     &       10          51 &$-$35          28  &    4$^d$  &&& 5.0 &1.25& 0.95\\
NGC  3524        &       11          06 &+11          23  &    4  &&SA II& 0.0 &1.21& 0.54\\
NGC  3564        &       11          10 &$-$37          33  &    4  &&&-2.0 &1.25& 0.37\\
              ESO  377- 29     &       11          14 &$-$33          54  &    4  &&&-2.0 &1.18& 0.40\\
NGC  3627     &          11          20 &+12          59&  5.1 && Sh & 3.0 &1.96 &0.34\\
NGC 3650 & 11 22 & +20 42 & 2 && SDB & 3.0 & 1.23 & 0.75 \\
              ESO  319- 26     &       11          30 &$-$41          03  &    4$^d$  &&& 5.3 &1.17& 1.02\\
NGC  3762        &       11          37 &+61          45  &    3$^f$  &&SA II, SDB& 1.0 &1.27& 0.58\\
NGC  3838        &       11          44 &+57          57  &    5.1  &&SA III&
-0.4 &1.19& 0.39\\
ESO  378- 20     &       11          47 &$-$37          33  & 5.1  &&&-2.1 &1.11& 0.33\\
              ESO  572- 44     &       12          01 &$-$20          29  &    1$^d$  &&J P, SDB& 3.0 &1.22& 0.77\\
IC    760        &       12          05 &$-$29          17  &    3  &3&J P&-1.7 &1.22& 0.53\\
              ESO  321- 10     &       12          11 &$-$38          32  &    1$^d$  &&SDB& 0.0 &1.27& 0.93\\
NGC  4215        &       12          15 &+06          24  &    3  &&&-1.0 &1.27& 0.42\\
              ESO  506-  3     &       12          21 &$-$25          04  &    3$^d$  &&& 2.3 &1.15& 0.48\\
              ESO  506- 33     &       12          40 &$-$25          19  &    4$^d$  &&&-2.0 &1.26& 0.43\\
              ESO  322- 51     &       12          40 &$-$41          36  &    3  &&&-3.0 &1.23& 0.54\\
NGC  4645B       &       12          43 &$-$41          21  &    4  &&&-2.2 &1.28& 0.42\\
ESO  322-100     &       12          49 &$-$41          27  &    2$^d$  &&&-2.0 &0.97& 0.37\\
              ESO  575- 59     &       13          07 &$-$19          23  &    4$^d$  &&&-0.8 &1.27& 0.66\\
NGC  5014        &       13          11 &+36          16  &    4  &4&SA II& 1.0 &1.23& 0.46\\
NGC  5193A       &       13          31 &$-$33          14  &    5.2$^d$
  &&SDB&-1.9 &1.09& 0.52\\
NGC  5253 & 13          39 & $-$31          38 & 4.0 &&SA III& 10.0 & 1.70 & 0.41\\
              ESO  445- 49     &       13          49 &$-$31          09  &    4$^d$  &&& 0.0 &1.15& 0.41\\
ESO 383- 85 & 13 49 &$-$33 05 & 3$^d$  &&& 0 $^6$ & 1.04$^6$ & 0.74$^6$ \\
ESO  383- 86     &       13          49 &$-$33          11  &    3$^d$ &&& -2 $^6$ & 1.00$^6$ & 0.70$^6$   \\
              ESO  445- 63     &  13 52  &  $-$30 49 &   1$^b$  &&J P& 5 $^5$ &  1.08$^6$ &  0.60$^6$  \\
NGC  5333        &       13          54 &$-$48          30  &    5.1  &&&-2.0 &1.28& 0.30\\
ESO  510- 13     &   13 55  & $-$26 46    &    3$^g$      &&& 1.0 & 1.29 & 0.17 \\
ESO  578- 25     &   14 08  & $-$20 00    &    2$^d$  &&& 1 $^6$ & 1.20$^6$ &0.60$^6$
    \\
          ESO  510- 74     &       14          08 &$-$26          56  &    3$^d$   &&&1.3 &1.23& 0.67\\
ESO 385- 8 & 14 19 & $-$34 51 & 4$^d$ &&& 7 $^6$ & 1.28$^6$ & 0.98$^6$ \\
    IC 4464               &       14          37 &-37    36  &    3   &&&-1.8 &1.21& 0.41\\
              ESO  581-  6     &       14          58 &$-$19          23  &    4$^d$   &&&7.0 &1.23& 0.86\\
              ESO  583-  8     &       15          57 &$-$22          29  &    4$^d$  &&J B& 6.0 &1.18& 0.97$^6$\\
              UGC 10205        &       16          06 &+30          06  &    3  & 3 && 1.0 &1.16& 0.23\\
              UGC 10535        &       16          46 &+06          28  &    4$^d$  &&& 2.0 &1.04& 0.76\\
ESO 71- 1 & 17 59 & $-$67 56 & 4$^d$ &&& 3 $^5$ & 0.99$^5$ & 0.70$^5$ \\

\end{tabular}
\end{center}
\end{table*}

\begin{table*}[hbtp]
Table \ref{listf} continued.
\begin{center}
\begin{tabular}{l|cccccrcc}
\multicolumn{1}{c|}{(1)} & (2) & (3) & (4) & (5) & (6) &
\multicolumn{1}{c}{(7)} & (8) & (9)  \\
\multicolumn{1}{c|}{Object} & RA & DEC & bulge & NIR & earlier &
\multicolumn{1}{c}{$T$} & log & log  \\
 & (2000) & (2000) & type & bulge & detections & & $D_{25}$ & $R_{25}$  \\
 & \,  & \, & \, & type & & \, & \, \\
\hline 
IC   4731        &       18          38 &$-$62          56  &    5.1  &&&-1.3 &1.17& 0.31\\
IC   4757        &       18          43 &$-$57          09  &    2$^b$ &&&-0.3 &1.14& 0.39\\
ESO  103- 59        &      18 45   &  $-$63 31  &   4  &&J B&   -2 $^6$ &1.14$^6$ & 0.85$^6$\\
IC   4767        &       18          47 &$-$63          24  &    1$^b$  &&J P, SDB&-1.0 &1.17& 0.51\\
IC   4808        &       19          01 &$-$45          18  &    5.1  &&& 5.0 &1.29& 0.37\\
NGC 6737 & \multicolumn{3}{c}{not a galaxy} & & SA I &  & &\\
              ESO  232- 21     &       19          44 &$-$51          36  &    4  &&&-1.5 &1.24& 0.74\\
ESO 398- 29 & 19 45 & $-$34 44 & 1$^b$ &&J P& 4 $^6$ & 1.23$^6$ & 0.89$^6$\\
ESO 105- 12 & 19 46 & $-$65 14 & 4 &&J P& -2 $^6$ & 1.08$^6$ & 0.78$^6$ \\
              ESO  461-  6     &       19          51 &$-$31          58  &    3$^d$  &&& 5.0 &1.21& 1.13\\
             ESO  339- 16     &  20 00       & $-$40 43 &    2$^d$  &&J P& 1 $^6$  & 1.15$^6$ &0.85$^6$ \\
              ESO  185- 53     &       20          03 &$-$55          56  &    2  &&J P, SDB&-2.0 &1.08& 0.42\\
IC   4937        &       20          05 &$-$56          15  &    1$^d$  &&J P, SDB& 3.0 &1.29& 0.64\\
              ESO  186- 36     &       20          21 &$-$53          45  &    2  &&& 1.0 &1.27& 0.55\\
ESO  528- 17      &      20          33 &$-$27 05 & 4$^d$ &&& 5.7 &1.20 &1.04 \\
              ESO  187-  8     &       20          43 &$-$56          12  &    5.5$^d$  &&& 6.0 &1.18& 0.88\\
              ESO  597- 36     &       20          48 &$-$19          50  &    1$^b$  &&J P, SDB&-1.7 &1.20& 0.68\\
IC   5053        &       20          53 &$-$71          08  &    4$^d$  &&& 1.0 &1.20& 0.43\\
              ESO  235- 35     &       21          01 &$-$52          00  &    2  &&J B& 3.0 &1.10& 0.76\\
ESO  466-  1     &    21    42 &$-$29    22  & 4$^d$ &&& 2.0 &1.14& 0.66\\
NGC  7123    &      21          50 &$-$70          20 &     4$^d$ &&SA II& 1.0&
1.48 & 0.32\\
ESO  189- 12     &       21          55 &$-$54          52  &    3$^d$
 &&SDB & 5.0 &1.22& 0.96\\
ESO 466- 46 & 22 02 & $-$31 59 & 5.1 &&J P& -2 $^5$ & 1.08 & 0.50 \\
IC   5199        &       22          19 &$-$37          32  &    3$^d$  &&J B, SDB& 3.0 &1.19& 0.85\\
IC   5267B       &       22          56 &$-$43          45  &    2$^d$ &&&-2.0 &1.27& 0.35\\
IC   5269        &       22          57 &$-$36          01  &    5.1  &&SDB&-1.8 &1.25& 0.34\\
NGC  7443        &       23          00 &$-$12          48  &    4  &&&-1.0 &1.17& 0.42\\
              ESO  604-  6     &       23          14 &$-$20          59  &    3$^d$  &&SDB& 4.0 &1.27& 0.85\\

\end{tabular}
\end{center}

Notes to the table: \\
\\
Col. (1) - (7):
s. Table \ref{listrc3} \\
\\
Col. (8): Decimal logarithm of the 25 ($B$) isophotal major axis
diameter from RC3, except
$^5$:
ESO-LV,  $^6$:  NED. \\
\\
Col. (9): Decimal logarithm of the 25 ($B$) isophotal
axis ratio ($a/b$) from RC3, except $^5$:
ESO-LV, $^6$: NED. \\

\end{table*}

\end{document}